\documentclass[prd,aps,amsmath,twocolumn,superscriptaddress,amssymb,preprintnumbers,reprint,nofootinbib]{revtex4-2}
\pdfoutput=1
\usepackage{graphicx,array}
\usepackage{hyperref}
\usepackage{color}
\usepackage{amsmath,amssymb,slashed,latexsym}
\usepackage{bm}
\usepackage{braket}
\usepackage{placeins}
\usepackage{adjustbox}
\usepackage{subcaption}
\usepackage{enumitem}
\usepackage{aas_macros}
\usepackage[dvipsnames]{xcolor}

\def\beq{\begin{equation}}
\def\eeq{\end{equation}}
\def\be{\begin{equation}}
\def\ee{\end{equation}}
\def\bea{\begin{eqnarray}}
\def\eea{\end{eqnarray}}

\usepackage{array}
\newcolumntype{P}[1]{>{\centering\arraybackslash}p{#1}}
\newcolumntype{M}[1]{>{\centering\arraybackslash}m{#1}}
\usepackage{tabularray}
\UseTblrLibrary{booktabs}

\definecolor{darkblue}{cmyk}{1,0.4,0,0.3}
\definecolor{violet}{cmyk}{0,1,0,0.2}
\hypersetup{colorlinks, bookmarksnumbered, citecolor=darkblue, linkcolor=darkblue, pdfstartview=FitH, urlcolor=darkblue, linktocpage}

\newcommand{\beqa}{\begin{eqnarray}}
\newcommand{\eeqa}{\end{eqnarray}}

\newcommand\q{{\bf q}}

\newcommand{\vk}{\mathbf{k}}

\newcommand{\vpsi}{\mathbf{\psi}}
\newcommand{\vp}{\mathbf{p}}
\renewcommand{\k}{\mathbf{k}} 

\def\nottilde#1{\overset{\ifx#1f\hspace{.5ex}\fi\lower0.8ex\hbox{\tiny$\nsim$}}{#1}}
\newcommand{\fnl}{f_{\rm NL}}

\def\hMpc{h{\text{Mpc}}^{-1}}

\begin{document}

\preprint{MIT-CTP/5956}
\preprint{KEK-Cosmo-0399}

\title{\boldmath Equilateral non-Gaussian Bias at the Field Level}

\author{Divij Sharma}
\email{d\_sharma@mit.edu}
\affiliation{Department of Physics and Kavli Institute for Astrophysics and Space Research,
Massachusetts Institute of Technology, Cambridge, MA 02139, USA}
\affiliation{The NSF AI Institute for Artificial Intelligence and Fundamental Interactions,
Cambridge, MA 02139, USA}

\author{James M. Sullivan}\thanks{Brinson Prize Fellow} 
\affiliation{Center for Theoretical Physics – a Leinweber Institute,
Massachusetts Institute of Technology, Cambridge, MA 02139, USA}

\author{Kazuyuki Akitsu}
\affiliation{Theory Center, Institute of Particle and Nuclear Studies,
High Energy Accelerator Research Organization (KEK), Tsukuba, Ibaraki 305-0801, Japan}

\author{Mikhail M. Ivanov}
\email{ivanov99@mit.edu}
\affiliation{Center for Theoretical Physics – a Leinweber Institute,
Massachusetts Institute of Technology, Cambridge, MA 02139, USA}
\affiliation{The NSF AI Institute for Artificial Intelligence and Fundamental Interactions,
Cambridge, MA 02139, USA}

\begin{abstract} 
 Primordial non-Gaussianity (PNG) is a common prediction of a wide class of inflationary models. Equilateral-type PNG, generically predicted by single-field inflationary models with higher-derivative interactions, imprints subtle but measurable signatures on the large-scale distribution of matter. 
 An important parameter of these imprints is the PNG-induced bias coefficient $b_\psi$, which quantifies how the abundance and clustering of 
dark matter halos 
and galaxies
respond to mode coupling in the initial conditions. Measuring $b_\psi$ is important for constraining equilateral PNG, yet it is notoriously challenging due to its degeneracy with Gaussian scale-dependent bias contributions. In this work, we present the first precision measurements of equilateral $b_\psi$ for dark matter halos using effective field theory
at the field level. 
We show that this approach  disentangles PNG effects from those of the Gaussian bias by virtue of noise variance 
cancellation. 
We compare our results 
with the phenomenological 
predictions based
on the Peak-Background 
Split model,
finding some agreement
at the qualitative level 
on the
redshift and mass dependence, but poor agreement 
at the 
quantitative level.  
We present a 
fitting formula for $b_\psi$
as a function of the linear bias, which
can be used to set priors 
in PNG searches 
with ongoing and future 
galaxy surveys.
\end{abstract}

\maketitle

\section{Introduction}\label{sec:intro}
The primordial accelerated expansion
of our Universe, known as cosmic inflation, 
is the leading theoretical framework
that addresses the origin and basic 
properties of cosmological fluctuations. 
Inflationary models generically predict
some small but observationally testable amount of primordial non-Gaussianity (PNG)
in the statistics of the cosmological 
initial conditions. This motivates a program of searching for these 
non-Gaussian signatures in the cosmic microwave background (CMB)
and large-scale structure (LSS) data
on the galaxy distribution. 
The extraction of PNG from LSS
has recently become a particularly 
important research topic spurred by 
the wealth of precision data from current and upcoming LSS surveys, such as 
DESI~\cite{Aghamousa:2016zmz}, Euclid~\cite{Laureijs:2011gra}, LSST~\cite{LSST:2008ijt}, and the Roman Space Telescope~\cite{Akeson:2019biv}.

The leading 
PNG signal is captured by the 
bispectrum, a Fourier image of the three-point correlation function 
of the primordial density
fluctuations. 
The PNG bispectrum 
is characterized by
its amplitude $\fnl$
and 
its shape
in momentum space~\cite{Babich:2004gb},
which encodes distinctive physical 
processes that could operate 
in the early Universe. 
For instance, the presence of multiple light fields
produces to the so-called local shape
of non-Gaussianity. This scenario
attracts significant attention
in the context of LSS, because local
PNG produces a scale-dependent modulation
of the galaxy power spectrum on large scales due to an enhanced correlation between large and small-scale structure. This effect, 
known as the scale-dependent bias~\cite{Dalal:2007cu}, has been
the leading probe 
used to set bounds on local PNG
with LSS datasets~\cite{Slosar08:png_obsv,MuellereBOSS_fnl,CastorinaeBOSS_fnl,LeistedtSDSSphoto_fnl,Cabass:2022ymb,DAmico:2022gki,Rezaie:2023lvi,DESI:2023duv,Barreira2022:fnl_boss,Cagliari:2023mkq,mccarthy_xcorr_cmb_fnl,Chaussidon:2024qni,Bermejo-Climent:2024bcb,Fabbian:2025fdk}. 
The scale-dependent local PNG bias
depends on a combination 
$\fnl^{\rm local} b_\phi$,
where $\fnl^{\rm local}$
is the amplitude of the skewness
of the primordial density distribution, 
and $b_\phi$
is the local PNG bias coefficient 
whose value depends on structure formation details. Strong constraints 
on local PNG are possible only 
if $b_\phi$ is known. 

Other important types of PNG are 
equilateral and orthogonal, which are common to single field
inflationary 
models with higher-derivative interactions. More fundamentally, 
equilateral non-Gaussianity 
is a consequence of the spontaneous breaking 
of de Sitter time translations by the inflaton in single-field inflation, which can be formalized in the context of the EFT of inflation~\cite{Cheung:2007st,Cheung:2007sv,Cabass2023PDU:eft}. This makes such PNG natural from the EFT viewpoint, 
though such a shape
can also appear in some multi-field models, e.g.~\cite{Tolley:2009fg,Green:2009ds}.
In what follows we focus on equilateral non-Gaussianity, though
much of our discussion applies to the 
orthogonal shape as well. 

Equilateral non-Gaussianity features a weak correlation between large and small scales, which suppresses its scale-dependent bias signature. 
Thus the leading observable used to constrain
this model is the bispectrum
of galaxies, which receives a
contribution from the primordial bispectrum~\cite{Cabass:2022epm,Cabass:2022epm,Philcox:2022frc,Chen:2024bdg,Ivanov:2024hgq}.
In actual 
cosmological searches of 
equilateral PNG with LSS data the scale-dependent bias is marginalized over~\cite{Cabass:2022epm,Chen:2024bdg,Ivanov:2024hgq}. 
This approach depends on priors on the
magnitude of the equilateral scale-dependent bias, which is poorly known. 
In principle, a detailed knowledge 
of the equilateral scale-dependent bias
may even help measure $\fnl$
from data, analogously to how the scale-dependent bias is used for this purpose 
in the context of the local PNG~\cite{Gleyzes:2016tdh}. 

The only estimates of the equilateral PNG
available until now are those based on the phenomenological
peak-background split (PBS) model~\cite{Schmidt:2010gw}. 
While this model provides  
relatively accurate predictions
for the scale-dependent halo bias
of local and 
quasi-single field PNG
types, see e.g.~\cite{2013PhRvD..88b3515S,Scoccimarro:2011pz,Biagetti:2016ywx,Desjacques:2016bnm, Goldstein:2024bky,Dalal:2007cu,desjacques_fnl_review,Assassi:2015fma,baldauf_gr,luisa_nn_bias,Lazeyras22,matarrese_lpng_bias,sefusatti_lpng_bias,pillepich_lpng_bias,dsi_lpng_bias,afshordi_lpng_bias,scoccimarro_lpng_bias,grossi_halo_lpng,jeong_komatsu_lpng_bias,pat_lpng_bias,giannantonio_lpng_bias,verde_09_lpng_bias,schmidt_kamionkowski_lpng_bias,djs_11_lpng_bias,kendrick_gnl_lpng_bias,quijote_png_halo,bareeria_lpng_h1,pat_bphi_08,adame_lpng_bias_unit,2024A&A...689A..69G,hadzhiyska_abacuspng,2025PhRvD.111j3521H,lazeyras_AB_quadratic_halos_21,2023MNRAS.524..325M,2017PhRvD..96h3528G,Sullivan:2024jxe,Sullivan2025:lpng_timev,Shiveshwarkar2025:lpng_su_ab,2011Desjacques:pngbias_pbs,dsi_lpng_bias,afshordi_lpng_bias,LoVerde2008:png_hmf},
it has not been 
tested for the case of the equilateral 
PNG. 
In addition,
precision measurements of
usual Gaussian bias parameters 
and higher derivative bias of DM halos 
\cite{Lazeyras:2019dcx,Ivanov:2024xgb,Ivanov:2025qie}
showed some departures from
the predictions based on the 
phenomenological models. 
The latter case is particularly concerning 
because the higher derivative bias
is conceptually similar to the 
case of the equilateral PNG bias.

Moreover, there is some difficulty
in the connection between the PBS model
predictions and the actual measurements
of these parameters
from the data. This difficulty arises
because the LSS data 
is commonly analyzed using 
the effective field theory (EFT)
approach~\cite{McDonald:2009dh,Baumann:2010tm,Desjacques:2016bnm,Ivanov:2022mrd},
where bias parameters are scheme and cutoff-dependent Wilson coefficients.
The PBS model allows one to 
predict bias parameters 
only in the 
specific limit when the EFT cutoff is set to zero,
which does not match the schemes used
in the EFT codes used in actual data
analyses, e.g. CLASS-PT~\cite{Ivanov:2019pdj,Chudaykin:2020aoj} or \texttt{velocileptors}~\cite{Chen:2020fxs,Chen:2020zjt,Chen:2021wdi}.\footnote{Some alternative codes are now available, incl. \textsc{PyBird}, \textsc{class one-loop}, \textsc{folps} and \textsc{pbj} \citep{DAmico:2020kxu,Linde:2024uzr,Noriega:2022nhf,Moretti:2023drg}.} 

Finally, the empirical halo-based
models are not expected to be accurate 
for galaxies, as exemplified 
by  
measurements of the local 
PNG
bias $b_\phi$ based on hydrodynamical
simulations~\cite{Barreira:2020ekm,Barreira:2021ukk,bareeria_lpng_h1,2022JCAP...01..033B,fondi_lpng_mt_forecast}.
This calls for a development 
of a systematic framework to accurately measure
the equilateral PNG biases 
from realistic simulations. 

These arguments motivate
a focused study of the equilateral PNG bias parameters
aimed at obtaining realistic priors
for actual data analyses.
In this work, we perform such a study
for the first time, focusing on the case of dark matter halos. 

From the practical point of view, 
this measurement is complicated by the 
fact that the equilateral PNG signature
is quite weak and degenerate with the usual Gaussian bias
contributions. Specifically, the scale-dependence
of the equilateral PNG interpolates between
a constant on large scales (degenerate with the linear bias) to a $k^2$ on small scales, 
degenerate with the higher-derivative bias.
In addition, measurements of $b_\psi$
from a limited range of wavenumbers 
are subject to degeneracies 
with other Gaussian bias contributions, 
e.g. from the cubic tidal operator 
$\Gamma_3$, whose low-$k$ limit 
is also proportional to a $k^2$-type bias~\cite{Assassi:2014fva}.
These degeneracies make it hard to 
robustly detect signatures
of the equilateral PNG bias. 
To overcome this difficulty, 
in this paper we employ the 
field-level EFT technique~\cite{Schmittfull:2018yuk,Schmittfull:2020trd} (see also refs.~\cite{Obuljen:2022cjo,Foreman:2024kzw,Schmittfull:2014tca,Lazeyras:2017hxw,
Abidi:2018eyd,
Schmidt:2018bkr,
Elsner:2019rql,
Cabass:2019lqx,
Modi:2019qbt,
Schmidt:2020tao,
Schmidt:2020viy,
Lazeyras:2021dar,
Stadler:2023hea,
Akitsu:2023eqa,
Nguyen:2024yth,
Akitsu:2024lyt,
Akitsu:2025boy}) which has been
previously used for precision
measurements of EFT parameters~\cite{Ivanov:2024hgq,Ivanov:2024xgb,Ivanov:2024dgv}, including
the local PNG bias
\cite{Sullivan:2024jxe}.
The key benefit of the field-level EFT
is the cosmic variance cancellation
by means of computing the EFT predictions 
for the exact initial conditions 
of the simulation. 
We show that the field-level comparison of Gaussian and non-Gaussian cosmologies offers a powerful way to isolate the equilateral PNG bias $b_{\psi}$ in simulations. 

This paper is organized as follows. Section~\ref{sec:model} reviews the halo bias model in the presence of PNG, including the relevant bias parameters and transfer functions for equilateral shapes. In Section~\ref{sec:simulations_and_fitting}, we describe our simulation setup with non-Gaussian initial conditions and our fitting procedure. In Section~\ref{sec:results}, we present the field-level measurements of halo bias due to equilateral PNG for various redshifts and compare them to theoretical predictions from the PBS formalism. We discuss the implications and robustness of our findings, and conclude in Section~\ref{sec:conclusions}.

\section{Theory \label{sec:model}}

In this Section, we present the theoretical framework used to model the imprint of equilateral-type primordial non-Gaussianity on the halo density field using EFT.

\subsection{Background}\label{sec:back}
PNG is a modification of the statistical properties of the assumed Gaussian initial conditions for structure formation. Many models of inflation predict interactions of the inflaton or additional light fields, which introduce couplings between long- and short-wavelength modes \cite{Chen:2010xka}. These nonlinearies generate a nonzero three-point function of the primordial Bardeen potential $\phi(\mathbf{k})$ \cite{1980Bardeen:phi}. As a result, the existence of a nonzero primordial bispectrum $B_\phi(k_1,k_2,k_3)$ is a signal of PNG and an important probe of non-standard models of inflation \cite{Bartolo_2004,Meerburg2019:png_white_paper}. The bispectrum is defined as
\begin{align}
\langle \phi(\mathbf{k}_1)\,\phi(\mathbf{k}_2)\,\phi(\mathbf{k}_3) \rangle
&\equiv B_\phi(k_1,k_2,k_3) (2\pi)^3\,\delta_D^{(3)}(\mathbf{k}_{123}),
\label{eq:bispectrum_def}
\end{align}
where $\mathbf{k}_{123} = \mathbf{k}_1 + \mathbf{k}_2 + \mathbf{k}_3$ and $\delta_D^{(3)}(x)$ is the three-dimensional Dirac delta function.
In what follows we will also use the primed
correlation functions
with the delta-function
stripped off, e.g.
\begin{align}
\langle \phi(\mathbf{k}_1)\,\phi(\mathbf{k}_2)\,\phi(\mathbf{k}_3) \rangle'
&= B_\phi(k_1,k_2,k_3) 
\label{eq:bispectrum_def_2}~\,.
\end{align}

Due to statistical homogeneity and isotropy, the bispectrum depends only on the magnitudes of three wavevectors $\mathbf{k}_1$, $\mathbf{k}_2$, and $\mathbf{k}_3$ that form a closed triangle. We define
\begin{equation}
    B_\phi(k_1, k_2, k_3) = 6 A_\phi^4 f_{\rm NL} \frac{\mathcal{S}(k_1, k_2, k_3)}{k_1^2 k_2^2 k_3^2} .
\end{equation}
where $A_\phi^2$ is the amplitude of the primordial power spectrum, 
\be 
\langle \phi(\k) \phi(\k')\rangle
=(2\pi)^3\delta_D^{(3)} (\k+\k')P_{\phi}(k,z)~\,,
\ee 
where 
$P_\phi(k) = A_\phi^2/k^3$, and $\mathcal{S}(k_1, k_2, k_3)$ encodes the shape of the bispectrum.

Different shapes of the bispectrum encode different physics during inflation \cite{chen_png_review}. For instance, a bispectrum that peaks in the squeezed triangle configuration ($k_1 \ll k_2 \approx k_3$) is a feature of multi-field inflationary models, where multiple light scalar fields contribute to the dynamics. 
This shape is called ``local'' PNG,
\begin{equation}
S_{\mathrm{local}}(k_1, k_2, k_3) = 
\frac{1}{3}\,\frac{k_1^2}{k_2 k_3} + 2~\mathrm{perms.} .
\end{equation}

Single-field models of inflation, where the inflaton is the only degree of freedom, give rise to bispectra which peak in equilateral ($k_1 \approx k_2 \approx k_3$) or flattened ($k_1 \approx k_2 \approx k_3/2$) triangles. 
These shapes define a type of ``non-local'' PNG. Such signatures can be described using a basis of two shapes - equilateral and orthogonal - parameterized by amplitudes $f_{\mathrm{NL}}^{\mathrm{equil}}$ and $f_{\mathrm{NL}}^{\mathrm{ortho}}$ \cite{Senatore:2009gt,Planck}. 
The equilateral and orthogonal shape templates are given by \cite{Senatore:2009gt, Babich:2004gb}
\begin{align}
S_{\mathrm{equil}}(k_1, k_2, k_3) 
&= \left( \frac{k_1}{k_2} + 5~\mathrm{perms.} \right) \notag\\
&\quad - \left( \frac{k_1^2}{k_2 k_3} + 2~\mathrm{perms.} \right) - 2 , \label{eq:Sequil}\\[6pt]
S_{\mathrm{ortho}}(k_1, k_2, k_3) 
&= (1 + p)\,\frac{\Delta}{e_3}
- p\,\frac{\Gamma^3}{e_3^2} , \label{eq:Sortho}
\end{align}
where $p = 8.52587$,
\begin{align}
\Delta &= (k_{123} - 2k_1)(k_{123} - 2k_2)(k_{123} - 2k_3), \nonumber \\[4pt]
k_{123} &= k_1 + k_2 + k_3, \quad 
e_2 = k_1 k_2 + k_2 k_3 + k_1 k_3, \nonumber \\[4pt]
e_3 &= k_1 k_2 k_3, \quad 
\Gamma = \frac{2}{3} e_2 - \frac{1}{3} (k_1^2 + k_2^2 + k_3^2) .
\end{align}
In this paper, we study the equilateral bispectrum shape to characterize its imprints on large-scale structure.

An important aspect of the 
bispectrum shape is its squeezed limit, 
where one momentum becomes soft.
The squeezed limit of the equilateral 
bispectrum reads, 
\be 
\begin{split}
  &  \lim_{k\to 0} B_{\phi}^{\mathrm{equil}}(k,q,|\k-\q|) \\
    & = 12 f_{\rm NL}^{\mathrm{equil}}\,\big(1-(\hat{\q}\cdot\hat{\k})^2\big)\,
   \frac{k^2}{q^2}\,P_\phi(k)\,P_\phi(q)~\,.
\label{eq:Squeezed_equil}
   \end{split}
\ee 
which can be contrasted with the local one 
\be 
\lim_{k\to 0} B_{\phi}^{\mathrm{local}}(k,q,|\k-\q|) = 4 f_{\rm NL}^{\mathrm{local}} P_\phi(k)\,P_\phi(q).
\ee 
The 
coupling between
small-scale and large-scale modes is suppressed
in the 
equilateral 
case by $k^2/q^2$,
which has profound consequences
for structure formation, 
and specifically, for a scale-dependent bias. 

To connect these primordial signatures to late-time observables, we must relate the primordial potential $\phi$ to the linear matter density field in cosmological perturbation theory $\delta_1$. Its power spectrum is given by
\be 
\langle \delta_1(\k,z) \delta_1(\k',z)\rangle
=(2\pi)^3\delta_D^{(3)} (\k+\k')P_{11}(k,z)~\,.
\ee 
Primordial non-Gaussianity
generates 
a non-zero three-point 
function of 
the $\delta_1$. 
To compute it, 
we use the 
transfer function $\mathcal{M}$, 
\begin{align}
\delta_{1}(\mathbf{k},z) &= \mathcal{M}(k,z)\phi(\mathbf{k}), \\
\mathcal{M}(k,z) &= \left(\frac{P_{11}(k,z)}{P_\phi(k)}\right)^{1/2}\,.
\label{eq:M_def}
\end{align}
For notational brevity, we will suppress the explicit $z$-dependence in the rest of the paper. The bispectrum of $\delta_1$ is then given by
\be 
\begin{split}
&\langle \delta_1(\k_1)\, \delta_1(\k_2)\, \delta_1(\k_3) \rangle'
= f_{\mathrm{NL}}\,B_{111}(k_1,k_2,k_3)\,, \\
&= \mathcal{M}(k_1)\,\mathcal{M}(k_2)\,\mathcal{M}(k_3)\,
B_{\phi}(k_1,k_2,k_3) \,,
\end{split}
\label{eq:B111}
\ee
where $B_{111}$ is the primordial contribution to the three-point function of the linear matter field, sourced by PNG.

To study the impact of PNG at an observable level, we look at the PNG effects on halo statistics. The starting point of studying how halos trace the underlying matter field is the bias expansion. First, for Gaussian initial conditions, consider the simplest Eulerian model with the linear bias $b_1$ to express the halo overdensity field $\delta_h$ as
\begin{align}
\delta_h(\k) = b_1 \, \delta(\k) + \epsilon(\k) \, ,
\label{eq:simple_model}
\end{align}
where $\delta$ is the nonlinear dark matter field, $\epsilon$ is a stochastic term that does not correlate with any perturbative field, and must be present since the relation between dark matter and halos is not perfectly deterministic. If the halo field $\delta_h$
is known, $b_1$ can be found by minimizing the mean-square difference 
$\langle|\delta_h - b_1 \delta|^2\rangle$, yielding
\begin{align}
b_1(k) = 
\frac{\langle \delta_h(\mathbf{k}) \, 
\delta^*(\mathbf{k}) \rangle}{\langle |\delta(\mathbf{k})|^2 \rangle} \, .
\label{eq:b1k}
\end{align}
The bias measured in this way is a function of $k$. To see how well the linear bias model works, we need to see the scales up to which $b_1(k)$ is a constant.

In the presence of PNG, short-scale statistics are modulated by long-wavelength perturbations, and additional terms are needed in the bias expansion for halo overdensity \cite{Desjacques:2016bnm, Assassi:2015fma}.
This effect is controlled by
the squeezed limit of the bispectrum. 
To account for it, 
one should introduce  
the initial gravitational
potential in the bias expansion, modulated
by the squeezed limit
momentum dependence. 
In particular, at leading order, equilateral PNG adds the field $\psi(\k) \propto k^2 \phi(\k)$ to the bias expansion with a bias parameter $b_{\psi}$ capturing the halo response to PNG \cite{Assassi:2015fma, Schmidt:2010gw},
\begin{align}
\delta_h(\k) = b_1 \, \delta(\k) + b_{\psi}f_{\rm NL}\psi(\k) + \epsilon(\k) \, .
\label{eq:simple_model_withPNG}
\end{align}
Using the transfer functions, 
this can be understood
as a scale dependence
of the linear bias
parameter. 
We can understand
it  by calculating the leading contribution to the halo-matter cross-correlation spectrum,
\begin{align}
P_{hm}(k) &\equiv \langle \delta_h(\mathbf{k}) \, \delta(\mathbf{k}) \rangle' \nonumber \\
&= b_1 \langle \delta_1(\mathbf{k}) \, \delta_1(\mathbf{k}) \rangle' + b_{\psi} f_{\mathrm{NL}} \langle \delta_1(\mathbf{k}) \, \psi(\mathbf{k}) \rangle' \nonumber \\
&= b_1 P_{11}(k) + b_{\psi} f_{\mathrm{NL}} P_{1\psi}(k) \nonumber \\
&= \bigl( b_1 + \Delta b(k) \bigr) P_{11}(k).
\label{eq:Phm}
\end{align}
where $P_{1\psi}(k)$ is the cross-correlation between the matter and the $\psi$ field. $\Delta b(k)$ above is the scale-dependent contribution introduced by $\psi$ to the linear bias,
\begin{align}
\Delta b(k) &\equiv f_{\mathrm{NL}} \, b_{\psi} \, 
\frac{(k/k_{\mathrm{NL}})^{2}}{\mathcal{M}(k)} ,
\label{eq:Delta_b}
\end{align}
where we introduced  the nonlinear scale of perturbation theory
$k_{\rm NL}$ 
for dimensional 
reasons. 

\begin{figure}
\includegraphics[width=1\columnwidth]{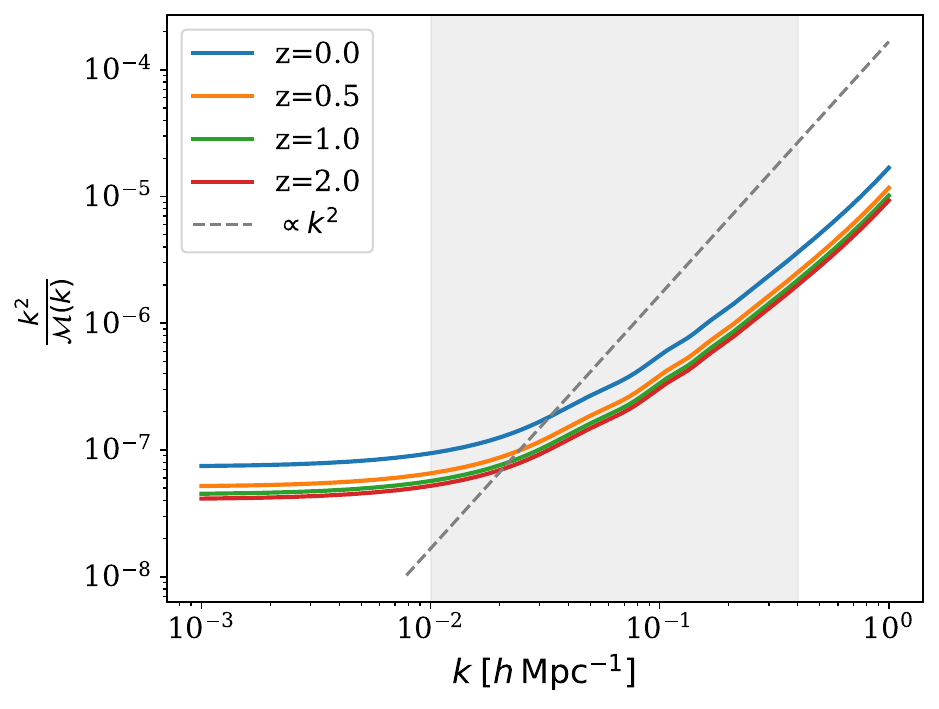}
\caption{Scale dependence of
the equilateral PNG bias
$k^{2}/\mathcal{M}(k)$ at multiple redshifts. The dashed line shows the asymptotic $k^2$ scaling. The shaded region ($0.01<$  $k/\,h\,\mathrm{Mpc}^{-1}<$  $0.4$ ) indicates the range probed by galaxy surveys.
}
\label{fig:k2_Over_M}
\end{figure}

We show the scale-dependent behavior of $\Delta b(k)$ across redshifts in Figure \ref{fig:k2_Over_M}. The shaded band indicates the range $0.01 < k/(h\,\mathrm{Mpc}^{-1})< 0.4$ that is most relevant for spectroscopic galaxy surveys. The curves show mild redshift dependence. On very large scales, the ratio flattens due to the $k^{-2}$ scaling of $\mathcal{M}(k)$ and behaves like a constant linear bias $b_1$, while at smaller scales, and in the shaded region, it asymptotically approaches a $k^2$ behavior \cite{Schmidt:2010gw,Desjacques:2011jb}. 

On small scales, 
the PNG-induced scale-dependent bias signature is somewhat degenerate with higher-derivative operators in the bias expansion 
$b_{\nabla^2\delta}\nabla^2\delta$~ \cite{Gleyzes:2016tdh,Assassi:2015fma,Desjacques:2016bnm,Assassi:2015jqa}. Importantly, while 
the PNG bias 
is fully degenerate with 
$b_1$
on large scales
and with $b_{\nabla^2\delta}$
on small scales, this degeneracy can be broken
in the intermediate transitional regime
which happens to match the 
range of scales
probed by galaxy surveys.

From the above discussion, it is natural to capture the PNG bias by promoting the bias parameters into scale-dependent transfer function $\beta_1(k)$ as is common in EFT forward models \cite{Schmittfull:2018yuk}. In this approach the $k$-dependence of $\beta_1(k)$
can be used to constrain the PNG-induced scale dependence and pin down $b_{\psi}$. With this motivation, we now turn to the full bias expansion at the one-loop order and then describe how we can use the transfer function approach to measure $b_{\psi}$.

\subsection{Halo bias expansion}

The basis for our EFT forward model~\cite{Schmittfull:2018yuk,Schmittfull:2020trd,Obuljen:2022cjo,Ivanov2024:PhRvD.110f3538I,misha_structure_eft_prior,Sullivan2025:mtng_highz,deBelsunce2025:lyal,cabass_field_level,Akitsu:2025boy,fabian_eft_like,fabian_s8_eft,julia_field_level_galaxies_24,Nguyen2024:FBI,Spezzati2025:fbi} is the bias expansion for the halo overdensity field $\delta_h$ relevant at one-loop order~\cite{Assassi:2014fva,Ivanov:2019pdj}. In a Universe with Gaussian initial conditions, this expansion reads
\begin{align}
\delta_h(\mathbf{k})\bigg|_{\rm n-pf}^{\rm EFT, G} &= 
b_1 \delta + \frac{b_2}{2} \delta^2 + b_{\mathcal{G}_2} \mathcal{G}_2 + b_{\Gamma_3} \Gamma_3 - b_{\nabla^2\delta} \nabla^2 \delta + \epsilon\,,\label{eq:eft_bias}
\end{align}
where we have emphasized that 
this is the model used to compute 
n-point correlation functions.
In the above equations, 
the standard perturbation theory kernels are defined as
\begin{align}
& \mathcal{G}_2(\mathbf{k}) 
= \int_{\mathbf{p}} F_{\mathcal{G}_2}(\mathbf{p}, \mathbf{k} - \mathbf{p}) 
   \delta(\mathbf{p}) \delta(\mathbf{k} - \mathbf{p})~,
   \label{eq:G2}\\
& F_{\mathcal{G}_2}(\mathbf{k}_1, \mathbf{k}_2) 
= \frac{(\mathbf{k}_1 \cdot \mathbf{k}_2)^2}{k_1^2 k_2^2} - 1\,,
\end{align}
where $ \int_{\mathbf{k}} \equiv \int \frac{d^3k}{(2\pi)^3}$.
The cubic Galileon operator $\Gamma_3$ is defined via the kernel
\begin{align}
F_{\Gamma_3} &= \frac{4}{7} \left( 1 - \frac{(\mathbf{k}_1 \cdot \mathbf{k}_2)^2}{k_1^2 k_2^2} \right) 
\left( \frac{((\mathbf{k}_1 + \mathbf{k}_2) \cdot \mathbf{k}_3)^2}{|\mathbf{k}_1 + \mathbf{k}_2|^2 k_3^2} - 1 \right). \label{eq:FGamma3}
\end{align}
The coefficients $b_1$, $b_2$, $b_{\mathcal{G}_2}$, $b_{\nabla^2\delta}$, and $b_{\Gamma_3}$ are free bias parameters for a given redshift and halo sample. 

We emphasize that $\delta_1$ used to construct the forward model is taken from the non-Gaussian initial conditions in the case of PNG simulations. It is also possible 
to use the Gaussian piece of $\delta_1$,
but the computation of transfer functions
becomes more involved in this case. 

As we have seen, PNG introduces an additional scale-dependent bias to the bias expansion given in the above eq.~\eqref{eq:eft_bias}~\cite{Assassi:2015jqa},
\begin{align}
\delta_h  (\mathbf{k})\bigg|_{\rm n-pf}^{\rm EFT, NG} &= \delta_h (\mathbf{k}) \bigg|_{\rm n-pf}^{\rm EFT, G}+ b_{\psi} f_{\mathrm{NL}} \left( \frac{k}{k_{\mathrm{NL}}} \right)^2\phi, 
\label{eq:eft_bias_NG}
\end{align}
where $k_{\mathrm{NL}} = 0.45\,h\,\mathrm{Mpc}^{-1}$ is the conventional normalization scale which roughly matches the nonlinear scale and we neglect higher order corrections \cite{Ivanov:2024xgb, Ivanov:2021kcd}.
$b_{\psi}$ in eq.~\eqref{eq:eft_bias_NG} is an additional bias parameter  of the model.

\subsection{Bias expansion at the field level} \label{subsec:TF_expansion}

In order to adsorb higher
order effects and 
minimize the dependence 
of the EFT forward model
on small scales the bias
parameters above are 
promoted to scale-dependent transfer functions \cite{Schmittfull:2018yuk}.

First, 
to account for bulk flows,  the halo overdensity operators are shifted by the Zel’dovich displacement field. These \emph{shifted operators} are defined as
\begin{equation}
\tilde{\mathcal{O}}(\mathbf{k}) = \int d^3q\, \mathcal{O}(\mathbf{q}) \exp\left[-i\mathbf{k} \cdot (\mathbf{q} + \vpsi_1(\mathbf{q}))\right],
\label{eq:shifted_op}
\end{equation}
where $\mathbf{q}$ denotes Lagrangian space (initial) coordinates, $\mathcal{O}(\mathbf{q})$ is a Lagrangian bias operator, and $\vpsi_1$ is the Zel’dovich displacement field given by
\begin{equation}
\vpsi_1(\mathbf{q}) = \int d^3k\, e^{i\mathbf{q} \cdot \mathbf{k}}\, \frac{i\mathbf{k}}{k^2} \delta_1(\mathbf{k}).
\label{eq:zel_displacement}
\end{equation}
It is important to note that
$\vpsi_1$
has non-Gaussian correlations.
To reduce degeneracies in the forward model description
and ensure numerical stability, the shifted operators are orthogonalized via the Gram–Schmidt process, ensuring
\begin{equation}
\langle \tilde{\mathcal{O}}_m^{\perp} \tilde{\mathcal{O}}_n^{\perp} \rangle = 0 \quad \text{if} \quad n \ne m.
\label{eq:gram_schmidt}
\end{equation}
The EFT forward model is built from the orthogonalized shifted operators
\begin{align}
\delta^{\mathrm{EFT}}_h(\mathbf{k}) &=  
\beta_1(k) \tilde{\delta}_1(\mathbf{k}) 
+ \beta_2(k) \left(\tilde{\delta}_1^2\right)^{\perp}(\mathbf{k}) \nonumber \\
&\quad + \beta_{\mathcal{G}_2}(k)\, \tilde{\mathcal{G}}_2^{\perp}(\mathbf{k}) 
+ \beta_3(k) \left(\tilde{\delta}_1^3\right)^{\perp}(\mathbf{k})\,,
\label{eq:eft_model_shifted}
\end{align}
where $\beta_i$
are scale-dependent 
transfer functions. 
In this approach, there is no need to introduce a separate set of transfer functions for the PNG operator $\psi(\mathbf{k})$; any additional $\beta_\psi(k)$ defined in this way would be degenerate with the existing $\beta_1(k)$.

Note that the above model
contains a single cubic
operator $\delta^3$. It is 
included in order to
ensure that the error power 
spectrum, defined as 
\be \label{eq:perr}
P_{\rm err}(k)=\langle |\delta^{\rm truth}_h(\k) - \delta_{h}^{\rm EFT}(\k)|^2\rangle'\,,
\ee 
is given by 
the stochastic contributions
at least up to the 3-loop
order~\cite{Schmittfull:2018yuk}, so that 
one can compute it in EFT as 
\be 
P_{\rm err}=\frac{1}{\bar n_h}\left(1+\alpha_0+\alpha_1 \left(\frac{k}{k_{\rm NL}}\right)^2+...\right)~\,,
\ee 
where $\bar n_h$
is the halo number density,
and $\alpha_{0,1}$
are stochastic EFT parameters
that capture the exclusion
effects~\cite{Baldauf:2013hka,sullivan_21,kokron_stoch_heft,Baldauf:2015fbu}. 

The transfer functions are fit using the halo overdensity field from simulations, $\delta^{\mathrm{truth}}_h(\mathbf{k})$,
 \begin{equation}
\beta_i(k) = \frac{\langle \mathcal{O}_i^{\perp *}(\mathbf{k})\, \delta_h^{\mathrm{truth}}(\mathbf{k}) \rangle'}{\langle |\mathcal{O}_i^{\perp}(\mathbf{k})|^2 \rangle'}.
\label{eq:transfer_function}
\end{equation}

\subsection{Transfer functions in EFT}

Before considering the effects of PNG, we recall the form of $\beta_1(k)$ in a Universe with Gaussian initial conditions~\cite{Ivanov:2024hgq}, which can be predicted in perturbation theory. We get, at the one-loop order \cite{Ivanov:2024xgb}
\begin{equation}
\begin{aligned}
\beta_1^{G}(k)
&= b_1 + b_{\nabla^2\delta} k^2
 + \frac{b_2}{2}\,
   \frac{\langle \tilde{\delta}_1 \tilde{\delta}_2 \rangle'}
        {\langle \tilde{\delta}_1 \tilde{\delta}_1 \rangle'}
 - b_1\,
   \frac{\langle \tilde{\delta}_1 \tilde{\mathcal{S}}_3 \rangle'}
        {\langle \tilde{\delta}_1 \tilde{\delta}_1 \rangle'} \\
&\quad
 + \left(b_{\mathcal{G}_2} + \frac{2b_1}{7}\right)
   \frac{\langle \tilde{\delta}_1 \tilde{\mathcal{G}}_2 \rangle'}
        {\langle \tilde{\delta}_1 \tilde{\delta}_1 \rangle'} \\
&\quad
 + \left(b_{\Gamma_{3}} + \frac{b_1}{6} + \frac{5}{2} b_{\mathcal{G}_2}\right)
   \frac{\langle \tilde{\delta}_1 \tilde{\Gamma}_3 \rangle'}
        {\langle \tilde{\delta}_1 \tilde{\delta}_1 \rangle'} ,
\end{aligned}
\label{eq:beta_1_G}
\end{equation}
where we use the superscript ``G'' to note that the derivation is carried out using Gaussian initial conditions. The $\tilde{\mathcal{S}}_3$ operator comes from the shift of the halo density field by the second-order displacement, $\mathcal{S}_3 \equiv \vpsi_2 \cdot \nabla \delta_1$.
In practice, the shifted 
correlators above are computed
using IR-resummed Eulerian
time-sliced perturbation
theory~\cite{Blas:2015qsi,Blas:2016sfa,Ivanov:2018gjr,Vasudevan:2019ewf,Ivanov:2024xgb}.

With PNG, we get additional terms due to the introduction of the $b_{\psi}$ term in eq.~\eqref{eq:eft_bias_NG} and the PNG-induced nonzero primordial bispectrum. When cross correlating equation
eq.~\eqref{eq:eft_bias_NG} with $\tilde{\delta}_1$, this leads to additional terms in eq.~\eqref{eq:beta_1_G}. We now derive these additional terms to get our model for $\beta_1(k)$ in the presence of PNG - $\beta_1^{\mathrm{NG}}(k)$.
First, we use the Eulerian kernel expansion for $\tilde{\delta}_1$ to express it in terms of $\delta_1$ as in \cite{Ivanov:2024xgb},
\begin{equation}
\tilde{\delta}_1 = \sum_{m=1}^{3} 
\left( \prod_{n=1}^{m} \int_{\k_n} \delta_1(\k_n) \right) 
(2\pi)^3 \delta_D^{(3)}\left(\k - \k_{1\ldots m} \right)\, \tilde{K}_n .
\label{tilde_delta_1}
\end{equation}
Explicitly, the three terms in the expansion are
\begin{align}
\tilde{\delta}_1^{(1)}(\vk) 
&= \tilde K_1(\vk)\,\delta_1(\vk), \\[6pt]
\tilde{\delta}_1^{(2)}(\vk) 
&= \int_{\vp} \tilde K_2(\vk-\vp,\vp)\,
    \delta_1(\vk-\vp)\,
    \delta_1(\vp), \\[6pt]
\tilde{\delta}_1^{(3)}(\vk) 
&= \int_{\vp_1,\vp_2} \tilde K_3(\vp_1,\vp_2,
       \vk-\vp_1-\vp_2) \nonumber \\
& \hspace{0.5cm}\quad\times \delta_1(\vp_1)\,
    \delta_1(\vp_2)\,
    \delta_1(\vk-\vp_1-\vp_2),
\end{align}
where the kernels $\tilde{K}_n$ are defined as
\begin{align}
\tilde{K}_1(\k) &= 1, \nonumber \\
\tilde{K}_2(\k_1, \k_2) &= 1 + \frac{1}{2} \left( \frac{\k_2 \cdot \k_1}{k_1^2} + \frac{\k_1 \cdot \k_2}{k_2^2} \right), \nonumber \\
\tilde{K}_3 &= \frac{1}{2} \frac{(\k_2 \cdot \k)(\k_3 \cdot \k)}{k_2^2\,k_3^2}. \label{eq:Kkernels}
\end{align}
The one-loop power spectrum of $\tilde{\delta}_1$ is given by
\begin{align}
\big\langle \tilde{\delta}_1(\mathbf{k})\,\tilde{\delta}_1(\mathbf{k}') \big\rangle'
\equiv P_{\tilde{1}\tilde{1}}(k) = P_{\tilde{1}\tilde{1}}^{G}(k) + P_{12}(k),
\label{eq:P11PNG}
\end{align}
where we have the usual expression for the Gaussian one-loop contribution
\begin{align}
P_{\tilde{1}\tilde{1}}^{G}(k) = P_{11}(k) + P_{22}(k) + P_{13}(k),
\label{eq:P11G}
\end{align}
with
\begin{align}
P_{11}(k)
&\equiv \big\langle \tilde{\delta}_1^{(1)}(\vk)\,
           \tilde{\delta}_1^{(1)}(\vk') \big\rangle' = \tilde K_1^2(\vk)\,P_{11}(k) = P_{11}(k), \\[6pt]
P_{22}(k)
&\equiv \big\langle \tilde{\delta}_1^{(2)}(\vk)\,
           \tilde{\delta}_1^{(2)}(\vk') \big\rangle' \nonumber \\
&= 2 \int_{\vp}
      \tilde K_2^{\,2}(\vk-\vp,\vp)\,
      P_{11}(|\vk-\vp|)\,P_{11}(p), \\[6pt]
P_{13}(k)
&\equiv \big\langle \tilde{\delta}_1^{(1)}(\vk)\,
           \tilde{\delta}_1^{(3)}(\vk') \big\rangle' \nonumber \\
&= 6\,\tilde K_1(\vk)\,P_{11}(k)
      \int_{\vp}\tilde K_3(\vp,-\vp,\vk)\,P_{11}(p), \\[6pt]
P_{12}(k)
&\equiv \big\langle \tilde{\delta}_1^{(2)}(\vk)\,
           \tilde{\delta}_1^{(1)}(\vk') \big\rangle' \nonumber \\
&= 2 f_{\rm NL}\int_{\vp}
      \tilde K_2(\vk-\vp,\vp)\,
      B_{111}\!\big(k,p,|\k-\vp|\big).
\end{align}

In eqs.~(\ref{eq:P11PNG},\ref{eq:P11G}), $P_{\tilde{1}\tilde{1}}^{G}$ is the one-loop power spectrum of the shifted field $\tilde{\delta_1}$ under Gaussian initial conditions, $P_{11}$ is the linear matter power spectrum, $P_{22}$ and $P_{13}$ are quadratic-quadratic and linear-cubic loop contributions respectively, and the $P_{12}$ term captures 
a mode-coupling contribution 
due to the primordial bispectrum~\cite{Assassi:2015jqa}. We compute the loop integrals in the above expressions with FFTLog~\cite{Simonovic:2017mhp}
using the CLASS-PT routine~\cite{Chudaykin:2020aoj}.

Now we can calculate $\beta_1^{\mathrm{NG}}(k)$.
We write the cross- and auto-spectra entering 
$\beta_1^{\rm NG}\equiv 
\langle \tilde\delta_1\,\delta_h^{\rm EFT}\rangle'
/\langle \tilde\delta_1\tilde\delta_1\rangle'$
as a Gaussian part plus a PNG correction, so the numerator and denominator are:
\begin{align}
\langle \tilde\delta_1(\k)\,\delta_h^{\rm EFT}(\k)\rangle'
&= P_{\tilde{1}\tilde{1}}^{G}(k) \beta_1^{\mathrm{G}}(k)  + f_{\rm NL}\,N(k),\\
\langle \tilde\delta_1(\k)\tilde\delta_1(\k)\rangle'
&= P_{\tilde{1}\tilde{1}}^{G}(k) \;+\; f_{\rm NL}\,D(k),
\end{align}
and, from eqs. (\ref{eq:eft_bias_NG}, \ref{tilde_delta_1}, \ref{eq:P11PNG}),
\be 
N(k) = N^{B_{111}}(k) + N^{b_{\psi}}(k)\,,\label{eq:N1}
\ee 
\be 
\begin{split}
N^{B_{111}}(k) &= \int_{\vp}\!\Bigg[\frac{b_2}{2}
+\Big(b_{\mathcal G_2}+\frac{2}{7}b_1\Big)F_{\mathcal G_2}(\k-\vp,\vp)\\
&\quad +\,2b_1\,\widetilde K_2(\k-\vp,\vp)\Bigg]\,
B_{111}\!\big(k,p,|\k-\vp|\big),\\
N^{b_{\psi}}(k) &= b_\psi\Big(\frac{k}{k_{\rm NL}}\Big)^2 \mathcal M^{-1}(k)\,
\langle \tilde\delta_1(\k)\tilde\delta_1(\k)\rangle', \\
f_{\rm NL}D(k) &= P_{12}(k)\label{eq:D1}.
\end{split}
\ee
The first term in eq.~\eqref{eq:N1} comes from the nonzero primordial bispectrum while cross correlating eqs. \eqref{eq:eft_bias_NG} and \eqref{tilde_delta_1} and the second term comes from the $b_{\psi}$ term in eq. \eqref{eq:eft_bias_NG}. Expanding to linear order in $f_{\rm NL}$,
\begin{align}
\beta_1^{\rm NG}(k)
&= \frac{P_{\tilde{1}\tilde{1}}^{G}(k) \beta_1^{\mathrm{G}}(k) + f_{\rm NL}N(k)}
{P_{\tilde 1\tilde 1}^{G}(k)+f_{\rm NL}D(k)} \notag\\
&= \beta_1^{\rm G}(k)
+ \underbrace{\frac{f_{\rm NL}}{P_{\tilde 1\tilde 1}^{G}(k)}\Big[N^{B_{111}}(k) - b_1 D(k)\Big]}_{\Delta\beta_1(k)} \notag\\
&\quad +\, b_\psi f_{\rm NL}\Big(\frac{k}{k_{\rm NL}}\Big)^2 \mathcal M^{-1}(k) + \mathcal O(f_{\rm NL}^2).
\end{align}
The $\Delta\beta_1(k)$ term\footnote{This is closely related to the coincidentally similarly named $\beta_m(k)$ presented e.g. in App. A of Ref.~\cite{2011Desjacques:pngbias_pbs} and discussed in Ref.~\cite{Biagettibphi}.} is given by
\begin{align}
\Delta \beta_1(k) 
&= \frac{f_{\mathrm{NL}}}{P_{\tilde{1}\tilde{1}}^{G}(k)}
   \int_{\vp} \Bigg[
      \frac{b_2}{2}
      + \Big(b_{\mathcal G_2}+\frac{2}{7}b_1\Big)
        F_{\mathcal G_2}(\k-\vp,\vp) \notag\\
&\quad\quad
      +\,2b_1\,\widetilde K_2(\k-\vp,\vp)
   \Bigg]\,
   B_{111}\!\big(k,p,|\k-\vp|\big) \notag\\
& - \frac{f_{\mathrm{NL}}b_1}{P_{\tilde{1}\tilde{1}}^{G}(k)}
   \int_{\vp} 2\widetilde K_2(\k-\vp,\vp)\,
   B_{111}\!\big(k,p,|\k-\vp|\big).
\label{eq:Delta_beta1_taylor_expanded}
\end{align}
And hence the $b_1 \widetilde K_2$ terms cancel.

So all in all, we have
\begin{align}
\beta_1^{\mathrm{NG}}(k) = \beta_1^{\mathrm{G}}(k) +  \Delta\beta_1(k) + b_\psi\, f_{\mathrm{NL}}\, \left( \frac{k}{k_{\mathrm{NL}}} \right)^2 \mathcal{M}^{-1}(k).
\label{eq:final_model}
\end{align}

The terms involved in eq.~\eqref{eq:Delta_beta1_taylor_expanded} are the same as in the cross-spectrum between the halo and matter density fields, which is already computed \textsc{CLASS-PT} \cite{Chudaykin:2020aoj,Cabass:2022epm}, so we can get the scale dependence in the above equation from the cross-spectrum,
\be 
\begin{split}
P_{hm,12}(k) 
&= \int_{\vp} \Big[
      2 b_1\, \widetilde K_2(\k-\vp,\vp)
      + \frac{b_2}{2} \\
&+ \Big(b_{\mathcal G_2}+\frac{2}{7}b_1\Big)\,
   F_{\mathcal G_2}(\k-\vp,\vp)\,\Big]
   B_{111}\!\big(k,p,|\k-\vp|\big).
\end{split}
\label{eq:Phm12}
\ee
The subscript ``12'' stand for the linear-quadratic contribution to the one-loop cross-spectrum, and $P^{b_1}_{hm,12}$, $P^{b_2}_{hm,12}$, $P^{b_{\mathcal{G}_2}}_{hm,12}$ are the contributions proportional to $b_1$, $b_2$ and $b_{\mathcal{G}_2}$ respectively.

The non-Gaussian
correction to $\beta_1$
can be re-written as
\begin{equation}
\begin{aligned}
\Delta \beta_1(k) 
&= \frac{f_{\mathrm{NL}}}{P_{\tilde{1}\tilde{1}}^{G}(k)}
   \Bigg[b_2 P^{b_2}_{hm,12}(k) 
      + \Big( b_{\mathcal G_2} + \frac{2}{7} b_1 \Big)\,
        P^{b_{\mathcal G_2}}_{hm,12}(k)
   \Bigg],
\end{aligned}
\label{eq:final_delta_beta1}
\end{equation}
with eq.~\eqref{eq:final_model} giving our model for $\beta_1^{\mathrm{NG}}(k)$.

As for the other transfer functions, since our EFT forward model uses shifted operators that are orthogonalized against $\tilde\delta_1$, so that by construction \(\langle (\tilde\delta_1^2)^{\perp}\,\tilde\delta_1\rangle' = 0\) and \(\langle \tilde{\mathcal G}_2^{\perp}\,\tilde\delta_1\rangle' = 0\), and when we project the halo field onto the orthogonal subspace to extract $\beta_2$ and $\beta_{\mathcal G_2}$, the $b_{\psi}$ term drops out:
\begin{align}
\big\langle \delta_h^{\mathrm{EFT}}\,(\tilde\delta_1^2)^{\perp}\big\rangle'
&\supset f_{\rm NL}\,b_\psi\,\Big(\frac{k}{k_{\rm NL}}\Big)^2\,\mathcal M^{-1}(k)\,
\underbrace{\big\langle \tilde\delta_1\,(\tilde\delta_1^2)^{\perp}\big\rangle'}_{=\,0}
\;=\; 0, \nonumber \\
\big\langle \delta_h^{\mathrm{EFT}}\,\tilde{\mathcal G}_2^{\perp}\big\rangle'
&\supset f_{\rm NL}\,b_\psi\,\Big(\frac{k}{k_{\rm NL}}\Big)^2\,\mathcal M^{-1}(k)\,
\underbrace{\big\langle \tilde\delta_1\,\tilde{\mathcal G}_2^{\perp}\big\rangle'}_{=\,0}
\;=\; 0.
\end{align}
Therefore, to linear order in $f_{\rm NL}$ and at the order we are interested in, $\beta_2(k)$ and $\beta_{\mathcal G_2}(k)$ are unaffected by $b_{\psi}$ and we get their usual form \cite{Ivanov:2024hgq}, 
\begin{align}
\beta_{2}(k) &= \frac{b_{2}}{2}, \label{eq:beta2_model} \\
\beta_{\mathcal{G}_2}(k) &= b_{\mathcal{G}_2} + \frac{2}{7} b_{1}. \label{eq:betaG2_model}
\end{align}

\section{Simulations and Fitting}
\label{sec:simulations_and_fitting}
In this Section, we describe the numerical simulations and fitting methodology that we use in our analysis to measure $b_{\psi}$. We describe the PNG simulations as well as other simulations used to quantify uncertainties in section \ref{sec:simulations}. We then outline our field-level fitting procedure, where transfer functions are extracted from our forward model and fitted to measure $b_{\psi}$, in section \ref{sec:fitting}.

\subsection{Simulations}
\label{sec:simulations}
We utilize a suite of four N-body simulations with equilateral PNG ran using the \textsc{Gadget-4} code \cite{Springel:2020plp}. 
As a control suite, we
also use four Gaussian simulations
with the same initial seeds 
as our PNG simulations. 

To amplify the PNG signal and practically measure the PNG bias \cite{Gleyzes:2016tdh}, these simulations are initialized with a large equilateral PNG amplitude of  $f_{\mathrm{NL}} = 1000$ with the modified version of \textsc{2LPTIC}~\cite{Crocce:2006ve, Scoccimarro:2011pz}. Each simulation evolves $1536^3$ dark matter particles in a periodic box of side length $L = 1\,h^{-1}\mathrm{Gpc}$, resulting in a particle mass of $\approx 2.4 \times 10^{10}\,h^{-1}M_\odot$. The cosmological parameters used for these simulations are: $h = 0.6766$, $A_s = 2.105 \times 10^{-9}$, $n_s = 0.9665$, $\Omega_m = 0.309$, $\Omega_b = 0.049$, $k_{\mathrm{pivot}} = 0.05\,\mathrm{Mpc}^{-1}$, $T_{\text{CMB}} = 2.725$ K. To identify dark matter halos, we use the \textsc{Rockstar} \cite{rockstar} phase-space halo finder. We use a variety of halo mass bins, and they are noted in section \ref{sec:results}. In the left panel of figure \ref{fig:field_level_fits}, we show the halo overdensity slice for one of our simulations at $z = 0.5$, using halos with masses log$M_h$ $\in$ [13.0, 13.5] $h^{-1 }M_{\odot}$. 

As a second dataset to compute uncertainties on our measurements, we use a large ensemble of $N$-body simulations from the Quijote suite \cite{Villaescusa-Navarro:2019bje} with Gaussian initial conditions. Each Quijote simulation evolves dark matter particles in a periodic box of $1~h^{-1}$Gpc on a side, with resolutions ranging from $256^3$ up to $1024^3$ particles per box. High-resolution (HR) Quijote runs used in this work have the finest resolution and evolve $1024^3$ particles. The simulations are run using the \textsc{Gadget-2} code \cite{Springel:2005mi}. Snapshots are saved at multiple redshifts ($ z = 0, 0.5, 1, 2, 3$), and we use $ z=0, 0.5, 1, 2$ in this work. The HR simulations correspond to an ensemble of 100 independent realizations and a fiducial cosmology consistent with Planck 2018 best-fit~\cite{Planck}. They share the same baseline cosmological parameters and volume, but have different initial random seeds.

\subsection{Extracting and fitting transfer functions}
\label{sec:fitting}
We use the field-level technique to measure the EFT parameters from our simulated halo catalogs. The field-level methodology enables cosmic variance cancellation, allowing for precision measurements of EFT parameters from our simulations. To compute the best-fit field-level EFT models, we use the publicly available perturbative forward model code \href{https://github.com/andrejobuljen/Hi-Fi_mocks}{\texttt{Hi-Fi mocks}}~\cite{Obuljen:2022cjo, Foreman:2024kzw,Ivanov:2024hgq,Ivanov:2024xgb,Ivanov:2024dgv}. 
\texttt{Hi-Fi mocks} computes the optimal field-level bias transfer functions by 
minimizing the error between the simulated
tracer 
density field and the EFT 
forward model, quantified 
by the error power spectrum~\eqref{eq:perr}.
Comparing these transfer functions
with the EFT computations allows one 
to determine the constant EFT parameters. 
In our implementation, the density and shifted fields are constructed on a regular Fourier grid with grid size $N_{\mathrm{mesh}} = 512$ for the four N-body PNG simulations and $N_{\mathrm{mesh}} = 256$ for the Gaussian Quijote HR simulations, which correspond to Nyquist wavenumbers $k_{\mathrm{Nyq}} \simeq 1.61\,h\,\mathrm{Mpc}^{-1}$ and 
$0.80\,h\,\mathrm{Mpc}^{-1}$, respectively.
These are much larger 
than scales 
used in our analysis. 

\begin{figure*}[htbp]
    \centering
    \includegraphics[width=0.7\linewidth]{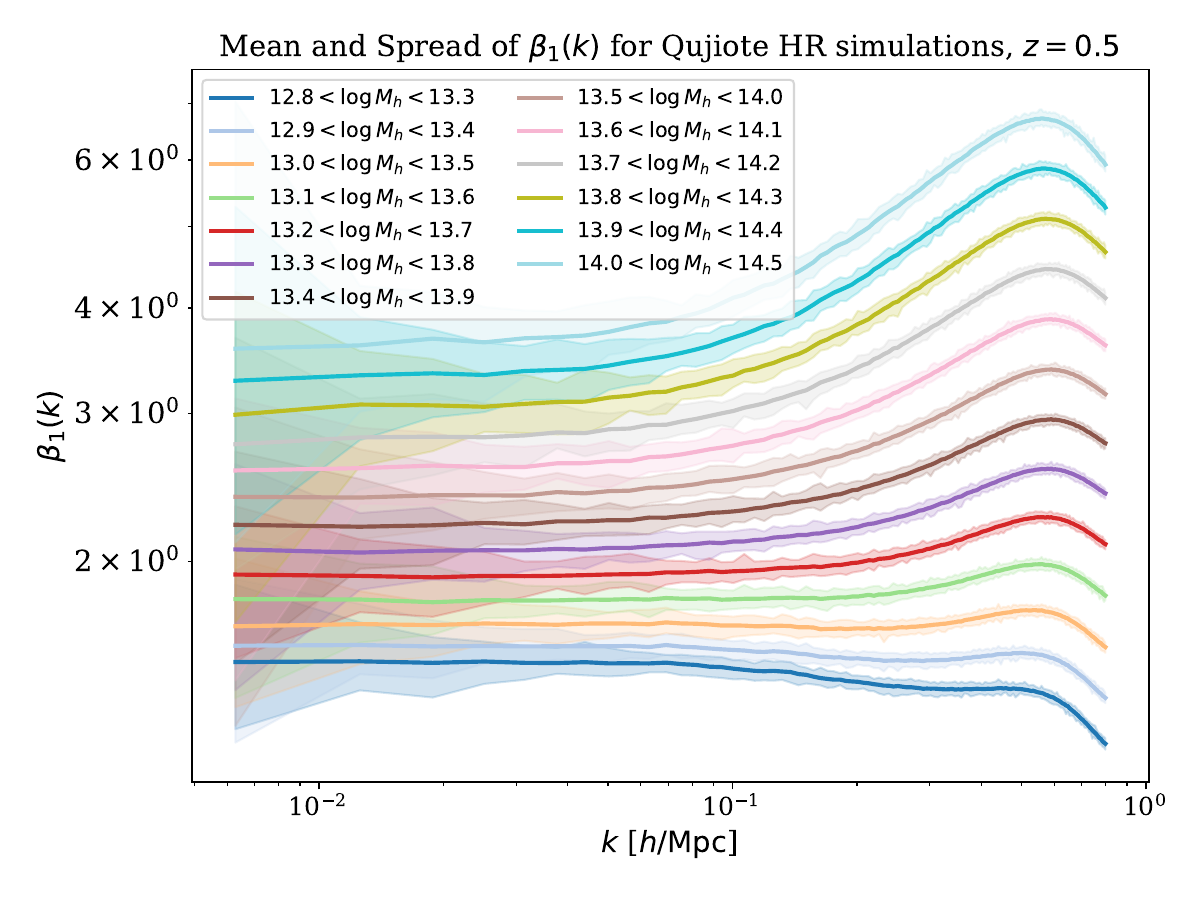}
    \caption{Mean and spread of the halo–matter transfer function $\beta_1(k)$ measured from the Gaussian Quijote HR simulations at $z = 0.5$. Each solid curve shows the mean $\beta_1(k)$ across 100 independent realizations for a given halo mass bin, in units of $h^{-1}M_{\odot}$, while the shaded region indicates the full range (minimum to maximum) across the ensemble. For a fixed mass bin, the transfer function is nearly scale-independent on large scales, but exhibits a strong upturn toward smaller scales ($k \gtrsim 0.1 \, h\,\mathrm{Mpc}^{-1}$), with the amplitude increasing systematically with halo mass.}
\label{fig:Quijote_spread}
\end{figure*}

To propagate uncertainties in our measurement of $\beta_X(k)$, $X \in \{1, 2, \mathcal{G}_2\}$, and compute error bars on our measured PNG bias $b_{\psi}$, we use the variance of $\beta_X(k)$ across the Quijote HR ensemble to build the covariance matrix.  Figure \ref{fig:Quijote_spread} shows the mean and spread of $\beta_1(k)$ across these realizations for different halo mass bins at $z=0.5$. For each mass bin, we compute the variance of $\beta_X(k)$ at fixed $k$ across the simulation ensemble to compute the covariance matrix for the transfer function. The diagonal elements of this matrix, $\sigma_{\beta_X}^2(k)$, are the effective scale-dependent errors on our measured $\beta_X(k)$. In our field-level fits using weighted least-squares, $\sigma_{\beta_X}(k)$ then quantifies the uncertainty vector in the $\chi^2$ minimization. Concretely, the function minimized by our weighted least-squares fits is the chi-square statistic written in terms of the likelihood $\mathcal{L}_{\beta_X}$ as
\begin{align}
-2 \ln \mathcal{L}_{\beta_X}(\theta) 
&= \sum_{k} 
\frac{\big[\beta_X^{\rm data}(k) - \beta_X^{\rm model}(k;\theta)\big]^2}
{\sigma_{\beta_X}^2(k)} \, ,
\label{eq:likelihood_betaX}
\end{align}
which corresponds to a Gaussian likelihood. Here, $\beta_X^{\rm data}(k)$ is the transfer function measured from our PNG simulations, $\beta_X^{\rm model}(k)$ are our models, and $\theta$ is the set of free bias parameters being fit. This way, our error bars on $b_{\psi}$ capture the stochastic scatter of $\beta_X(k)$ due to cosmic variance and shot noise.

As in \cite{Ivanov:2024jtl}, we use the following template to fit and extract the bias parameters $b_2$, $b_{\mathcal G_2}$ from $\beta_{2}, \beta_{\mathcal{G}_2}$:
\begin{equation}
\beta_X^{\rm model}(k;\{a^{X}_{0}, a^{X}_{1}, a^{X}_{2}\}) = a^{X}_{0} + a^{X}_{1}\,k^{2} + a^{X}_{2}\,k^{4},
\label{eq:beta_template}
\end{equation}
for $X = \{2, \mathcal{G}_2\}$.
Then we take $a^{X}_{0}$ as a measurement of the corresponding
bias parameter. We always fix $b_2$ and $b_{\mathcal G_2}$ to the transfer function best-fit values. 

As for $b_1, b_{\Gamma_3}, b_{\nabla^2 \delta}, b_{\psi}$, we fit them from the shape of $\beta_1(k)$ using eq.~\eqref{eq:final_model} in three different ways to get $b_{\psi}$ measurements:

\paragraph{All parameters free.}
First, in our analysis, the simplest way to start is by fitting $\theta=\{b_1, b_2$, $b_{\mathcal G_2}, b_{\Gamma_3}, b_{\nabla^2 \delta}, b_{\psi}\}$ to $\beta_1^{\rm model}(k)=\beta_1^{\rm NG}(k)$ (eq.~\eqref{eq:final_model}).

However, since this leads to many free parameters with known degeneracies, it would be a good strategy to reduce the number of free parameters during our fits. 

\paragraph{Fixing 
$b_{\Gamma_3}, b_{\nabla^2\delta}$ to their Gaussian values.}
One way to reduce the number of free bias parameters is to fix them to their best-fit Gaussian values extracted from the Gaussian control simulations. To that end we first fit $\theta=\{b_{\Gamma_3}, b_{\nabla^2 \delta}\}$ to $\beta_1^{\rm model}(k)=\beta_1^{\rm G}(k)$ (eq.~\eqref{eq:beta_1_G}) and then use these values when fitting $\theta=\{b_1, b_2$, $b_{\mathcal G_2}, b_{\psi}\}$ to $\beta_1^{\rm model}(k)=\beta_1^{\rm NG}(k)$. 

The above approach, however, 
has an important limitation. 
The usual bias parameters
of PNG simulations depend
on $f_{\rm NL}$.
In the context of 
simple analytic models
like PBS, this 
follows from
the rescaling 
of the peak height 
with $\fnl$~\cite{Valageas:2009vn,2011Desjacques:pngbias_pbs,Biagettibphi,afshordi_lpng_bias}.
\be 
\nu^* = \frac{\delta_c}{\sigma(M)}\left(1-\frac{\delta_c}{3}S_3(M)\right)^{1/2}~\,,
\ee 
where $\delta_c$
is the spherical collapse critical density, 
$\sigma(M)$
is the mass variance
of the initial density field
in a spherical
cell enclosing 
mass $M$, and $S_3$
is the $\fnl$-dependent skewness
of the PDF of the initial
density.

Let us present now a simple
EFT-based argument
that can be used 
to estimate this 
effect entirely 
within the 
effective description. 
In principle,
the EFT expansion should 
include cubic terms like 
\be 
\delta\big|^{\rm n-pf}_{\rm EFT, NG}\supset \fnl \psi \delta^2_1~\,.
\ee 
The contribution from this term
is fully absorbed 
into $b_1$ and $b_\psi$ at 
halo power spectrum level to  one-loop order.
Let us use the EFT naturalness
arguments to estimate the typical shift of $b_1$
due to this effect. We have 
\be 
P^{\psi \delta^2_1}_{13}\supset  3b_1\fnl P_{11}(k)\int_\q P_{1\psi}(q)\equiv 2 b_1 \Delta b_1 \big|_{\rm \psi \delta^2_1} P_{11}(k)~\,.
\ee 
A direct evaluation in our fiducial cosmology at $z=0.5$ gives
\be 
\label{eq:est_shift}
\begin{split}
\Delta b_1 \big|_{\rm \psi \delta^2_1} \simeq & \frac{3}{2}\fnl \int_{q\leq \Lambda}\frac{q^2dq}{(2\pi)^3} \frac{q^2}{k^2_{\rm NL}}\frac{P_{11}(q)}{\mathcal{M}(q)}\\
\lesssim & ~0.1\left(\frac{\fnl}{10^3}\right)~\,,\quad \text{for}\quad \Lambda=0.7~h\text{Mpc}^{-1}\,.
\end{split}
\ee 
One can obtain similar estimates for the shifts of the 
other bias parameters
following the arguments on their
renormalization at the level
of the one-loop bispectrum~\cite{Assassi:2014fva}.
While
the renormalization of 
Gaussian bias parameters
due to $\fnl$ is a very weak effect, its magnitude 
is comparable with the 
physical effect of the scale-dependent bias, and hence
re-using the Gaussian
values for these parameters
may not be a good approximation.
Therefore, this option is considered to be the most extreme
in our list. 

\begin{figure}[htbp]
\centering
\includegraphics[width=\columnwidth]{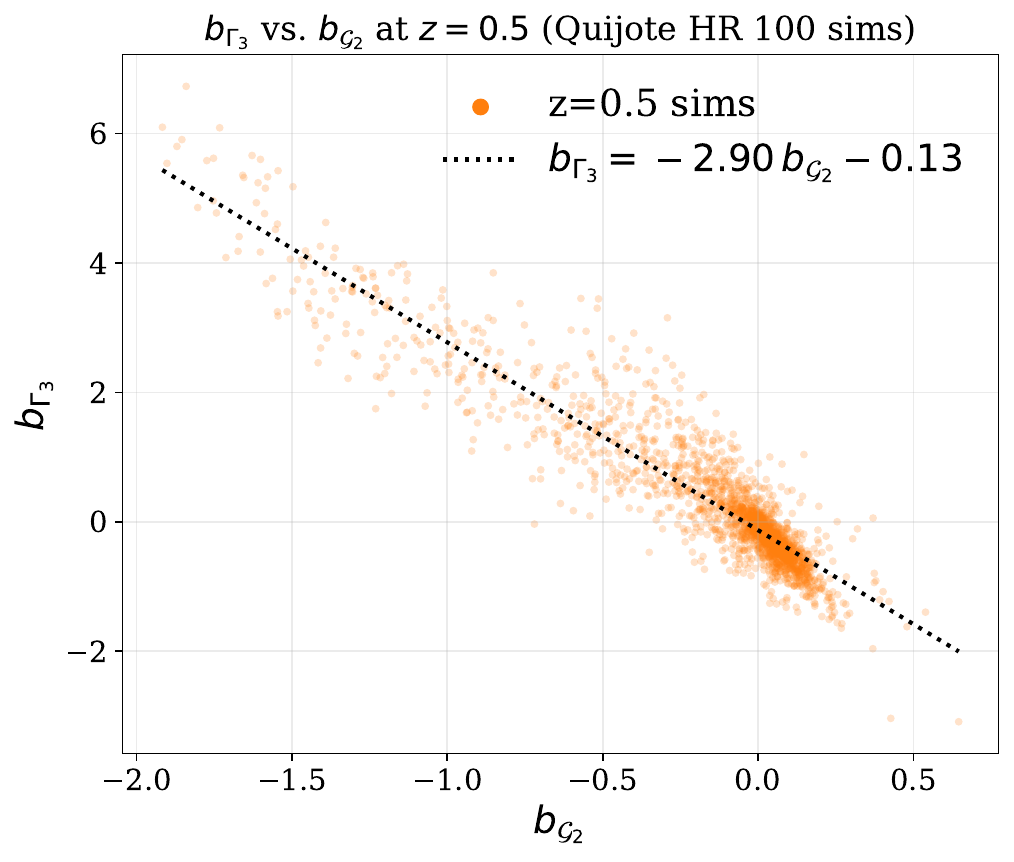}
\caption{Halo bias measurements of $b_{\Gamma_3}$ versus $b_{\mathcal{G}_2}$ from 100 Quijote HR simulations with Gaussian initial conditions at $z=0.5$. Each point corresponds to a fit for one simulation and halo mass bin. The black dotted line shows the empirical relation $b_{\Gamma_3} = -2.90b_{\mathcal{G}_2} - 0.13$. The measurements closely follow the linear relation across the full range of $b_{\mathcal{G}_2}$ values.}
\label{fig:bGamma3_vs_bG2}
\end{figure}

\paragraph{Using an empirical relation for $b_{\Gamma_3}$, with all other parameters free.}
In another approach to reduce the number of free parameters, we utilized a tight empirical relation between $b_{\Gamma_3}$ and $b_{\mathcal{G}_2}$ in order to fix $b_{\Gamma_3}$ using our measured values of $b_{\mathcal{G}_2}$. This
tight empirical relation has been noticed before in~\cite{Ivanov:2024hgq} in the context of galaxies.  
To validate it, we directly measure both $b_{\mathcal{G}_2}$ and $b_{\Gamma_3}$ in the Quijote HR Gaussian simulations. Figure~\ref{fig:bGamma3_vs_bG2} shows the resulting measurements at $z=0.5$, where each point corresponds to a halo mass bin in one of the 100 realizations. We see a tight correlation between $b_{\Gamma_3}$ and $b_{\mathcal{G}_2}$, similar to the one found by~\cite{Ivanov:2024hgq}\footnote{Here we use a more conservative scale cut $k_{\rm max}=0.2~\hMpc$ which leads to a relation slightly 
different from the one found in~\cite{Ivanov:2024hgq}
at $k_{\rm max}=0.4~\hMpc$.},
\[ 
b_{\Gamma_3} = -2.90\, b_{\mathcal{G}_2} - 0.13\,,\]
confirming that this empirical formula provides an accurate description of $b_{\Gamma_3}$ across halo populations. 

While this correlation has been
derived in the context of the Gaussian simulations, 
is expected to hold even in PNG universes. 
The reason is that 
under the assumption of a universal mass function, 
at the leading approximation the 
PNG simply rescales the peak height without
modifying the shape of the halo mass function~\cite{Valageas:2009vn}.
Then it is easy to show that 
local 
halo bias parameters $b_n$ in PNG Universes retain the same dependency on the (modified) peak 
height as in the Gaussian universe.
This implies that correlations between  
local bias parameters do not depend on PNG in this approximation.

As far as the non-local bias 
parameters such as $b_{\mathcal{G}_2}$
and $b_{\Gamma_3}$
are concerned, empirically
it was also found that they
primarily depend 
on the peak height just as the local biases~\cite{Ivanov:2024xgb,Ivanov:2025qie}\footnote{This simplified argument neglects the impact of the tidal field on the bias parameters through an ellipsoidal collapse barrier \cite{Desjacques:2016bnm}.}.
Thus, it is natural 
to extend the above argument
to these parameters as well. 
In other words,
there is a generalized 
definition of 
the peak height 
in the PNG case
that implies a degeneracy between $\fnl$
and the halo mass.
This degeneracy is very similar to the degeneracy between the amplitude of mass fluctuations 
and the halo mass that can 
be used to explain
the approximate 
cosmology and 
redshift-independence of the correlations between Gaussian 
halo
bias parameters~\cite{Ivanov:2024xgb,Ivanov:2025qie,Desjacques:2016bnm,Tinker:2008ff}.

While the above argument
relies on certain simplifying assumptions and extrapolations,
it suggests that 
one can treat 
the correlations between EFT parameters
as approximately $\fnl$ independent.
In agreement with this, we will
see shortly that the actual
Gaussian bias parameters $b_2$
and $b_{\mathcal{G}_2}$
as functions of $b_1$
from the simulations 
do not appreciably depend
on $\fnl$ for $b_1\lesssim 4$.

To sum up, as final fitting strategy, we use this relation to fix $b_{\Gamma_3}$ when fitting $\theta=\{b_1, b_2$, $b_{\mathcal G_2}, b_{\nabla^2 \delta}, b_{\psi}\}$ to $\beta_1^{\rm model}(k)=\beta_1^{\rm NG}(k)$.

\section{Results \label{sec:results}}
In this Section, we present the results using our field-level methodology to measure $b_{\psi}$. We show that a field–level EFT forward model with scale–dependent transfer functions accurately reconstructs the simulated halo field, leaving only small–scale residuals (section \ref{sec:overview}). Measuring the transfer functions shows a systematic PNG-induced offset relative to Gaussian initial conditions, allowing for a nonzero measurement of $b_{\psi}$. We make these measurements of $b_{\psi}$ in section \ref{sec:fit_results} using the three fitting strategies discussed in section \ref{sec:simulations_and_fitting} and extend our analysis to other redshifts in section \ref{sec:other_redshift_results}. We also compare our measurements against theoretical predictions from the PBS framework in section \ref{sec:PBS}, and find that $b_\psi$ strengthens with halo mass and redshift, in qualitative agreement with PBS expectations.

\subsection{Overview of fields and transfer functions}
\label{sec:overview}
\begin{figure*}[htbp]
    \centering
    \includegraphics[width=1.0\textwidth]{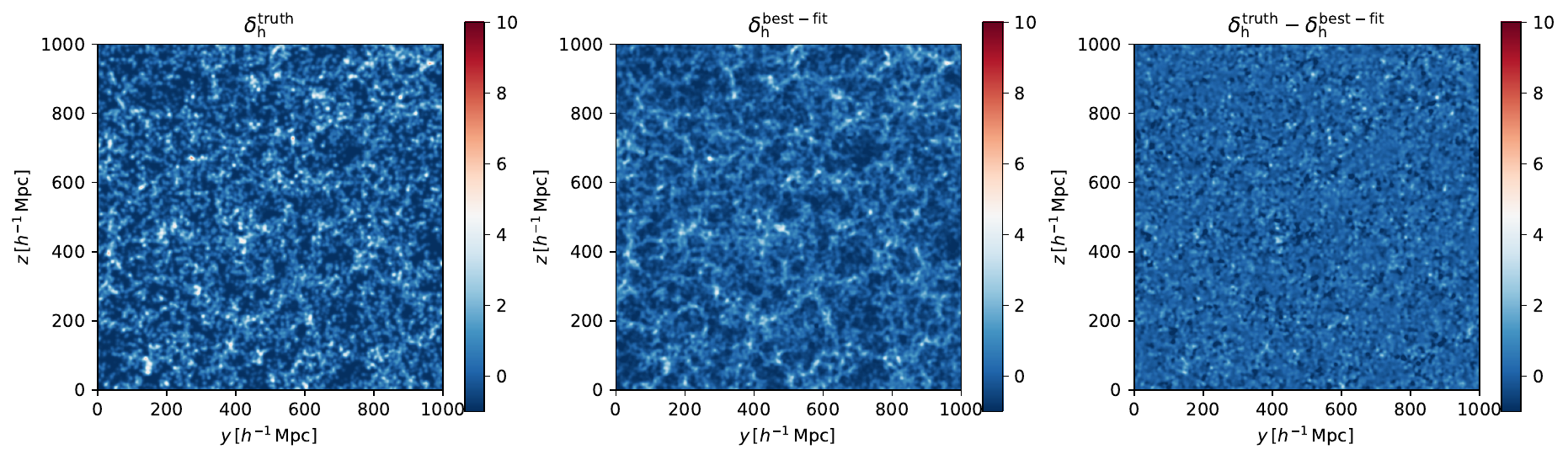}
    \includegraphics[width=1.0\textwidth]{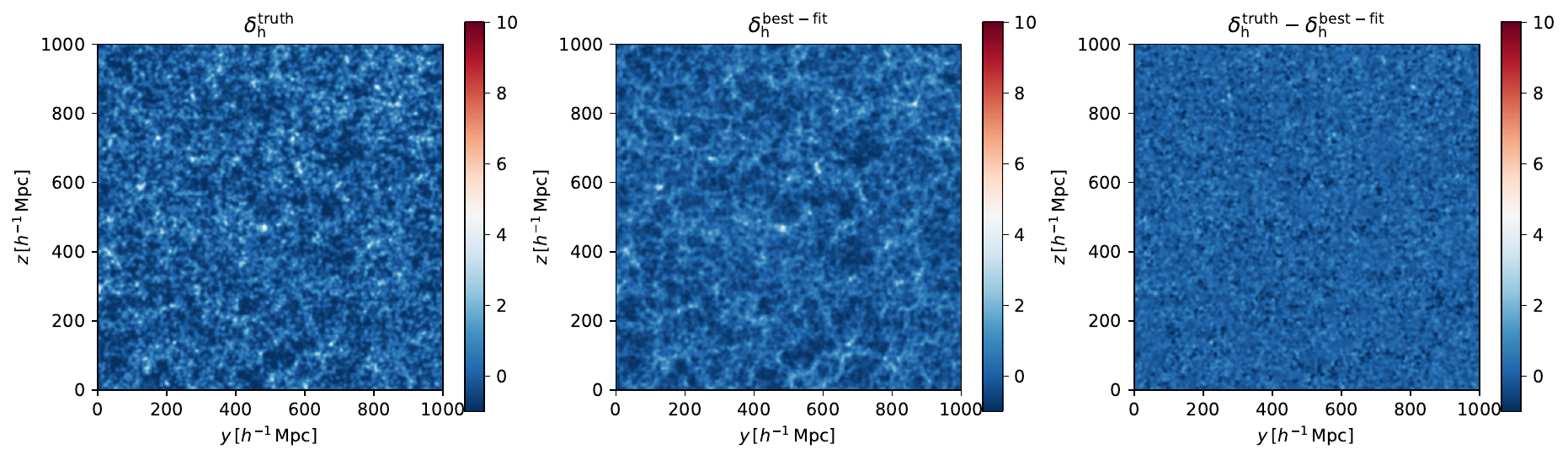}
    \caption{Field-level comparison between the simulated halo overdensity field and the best-fit forward model for $z=0.5$ and halo masses $\log M_h$ $\in$ [13.0, 13.5] $h^{-1}M_{\odot}$ computed using \texttt{Hi-Fi mocks}. The top row shows the comparison for a simulation with PNG, while the bottom row presents the corresponding comparison for the Gaussian (PNG-free) initial conditions. The left panel shows the simulated halo overdensity slice, $\delta_h^{\mathrm{truth}}$, taken directly from the simulation. The middle panel displays the corresponding best-fit forward-modeled field, $\delta_h^{\mathrm{best-fit}}$, computed by linearly combining the basis operators with the optimal transfer functions. The right panel shows the residual field, $\delta_h^{\mathrm{truth}} - \delta_h^{\mathrm{best-fit}}$, which quantifies the model error. }
    \label{fig:field_level_fits}
\end{figure*}

In Figure \ref{fig:field_level_fits} we show the field-level comparison, with and without PNG, between the simulated halo overdensity field and the best-fit forward model for $z=0.5$ and halo masses log$M_h$ $\in$ [13.0, 13.5] $h^{-1 }M_{\odot}$. The left panels show slices of the simulated field, while the middle panels show the corresponding best-fit forward model. The slices shown here correspond to a finite slice depth of 120 $h^{-1}{\rm Mpc}$, and have been smoothed with a Gaussian kernel of width $R=4\,h^{-1}{\rm Mpc}$ in order to highlight the large-scale modes relevant to our EFT description. Visually, the prominent structures such as clusters, filaments, and voids are well matched between the fields. The similarity between the left and middle panels shows that the EFT expansion accurately reproduces the large scale features of the simulated field. This is true for simulations with PNG (top row) as well as Gaussian initial conditions (bottom row). Comparing the two simulation halo fields, we see that both exhibit the broad cosmic web features, but the PNG simulation shows more clustering around overdense regions, as expected because of the large $f_{\rm NL}$ value used in our PNG simulations. The rightmost panel shows the residual field $\delta_h^{\rm truth}-\delta_h^{\rm best-fit}$, which is dominated by small-scale fluctuations that are not captured by our model. Overall, the forward model with the optimized transfer functions can accurately reconstruct the halo overdensity field. 

\begin{figure*}[htbp]
    \centering
    \includegraphics[width=\textwidth]{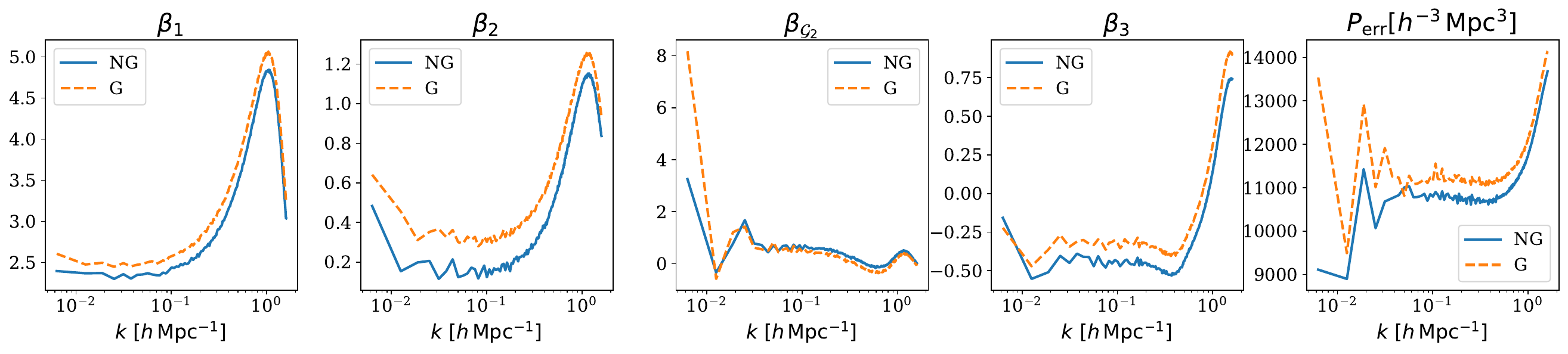}
    \caption{Comparison of the field-level bias transfer functions $\beta_1(k), \beta_2(k), \beta_{{\mathcal{G}}_2}(k), \beta_3(k)$, along with the residual power spectrum, $P_{\rm err}(k) \equiv \big\langle \, \big| \delta^{\rm best-fit}_h(\mathbf{k}) - \delta^{\rm truth}_h(\mathbf{k}) \big|^2 \big\rangle'$, for halo masses $\log M_h$ $\in$ [13.5, 14.0] $h^{-1 }M_\odot$ at redshift $z = 0.5$. Results are averaged across the four simulations and shown with (blue) and without (orange) PNG, measured using \texttt{Hi-Fi mocks}. The scale-dependent offset induced by equilateral PNG is visible in the PNG curve.}
    \label{fig:transfer_funcs}
\end{figure*}

To quantify how the halo overdensity field traces the underlying matter density field, we measure the EFT transfer functions, described in section \ref{subsec:TF_expansion}, in simulations with Gaussian initial conditions as well as with PNG. These transfer functions encapsulate the scale dependence of the relation between halos and matter, and they are the key observables for distinguishing Gaussian from non-Gaussian initial conditions in our field-level approach. These measured transfer functions are shown in figure \ref{fig:transfer_funcs} for $z=0.5$ and halo masses $\log M_h \in [13.5, 14.0]\, h^{-1 }M_\odot$. The orange curves are measurements from simulations with Gaussian initial conditions, while the blue curves are from simulations with equilateral PNG. The Gaussian transfer functions are roughly constant on large scales but show a scale dependence at smaller scales due to nonlinear bias effects. When equilateral PNG is introduced, the transfer functions are shifted relative to the Gaussian case.
The magnitude of these shifts
is consistent with our theoretical estimate from eq.~\eqref{eq:est_shift}.

We see that almost all of the measured transfer functions in the PNG simulations lie systematically below their Gaussian counterparts, with the exception of $\beta_{\mathcal{G}_2}$, showing that halos are less biased in our $\fnl=1000$ simulations. We observe a clear offset at all scales, consistent with the theoretical expectation \cite{Schmidt:2010gw} that equilateral PNG produces a scale-dependent modification to halo clustering that is more subtle than the case of local PNG, which induces a bias $\propto k^{-2}$. The clear offsets show that the field-level approach is sensitive to the PNG-induced bias differences. 

The right panel of Fig.~\ref{fig:transfer_funcs} also shows the error power spectrum of the field level model with respect to the simulated halos.
In the leading approximation, the large-scale amplitude of the power spectrum is given by the Poisson expectation.
PNG modulates the total number density of halos of a given mass (e.g. through the change in the mass variance) - more positive $\fnl$ generally leads to larger (high-mass) halos (see e.g. Fig.~1 of Ref.~\cite{quijote_png_hmf}).
Since the linear bias of a higher number density sample of halos selected by mass is smaller, this is consistent with the reduction in $b_1$ in the non-Gaussian simulations seen in the left panel of Fig.~\ref{fig:transfer_funcs}.
Additional PNG-dependent changes in the scale dependence of $P_{\mathrm{err}}$ are due e.g. to additional stochastic terms correlated with the potential (starting with $\epsilon_\psi \psi$ \cite{Desjacques:2016bnm}).

\subsection{Field-level bias measurements}\label{sec:fit_results}

The offset observed in the field-level transfer function $\beta_1(k)$ between simulations with Gaussian initial conditions and PNG allows us to measure a nonzero value for the PNG bias parameter $b_{\psi}$ by using eq.~\eqref{eq:final_model}. The strength of the PNG bias signal in our field-level fits depends on the range of scales included, as well as the treatment of higher-order bias parameters. Higher-order bias terms, such as $b_{\Gamma_3}$ and $b_{\nabla^2 \delta}$ in eq.~\eqref{eq:beta_1_G} absorb part of the small-scale dependence of $\beta_1(k)$. Since the fit results are sensitive to the amount of small-scale information we include as well as how we treat higher-order and higher-derivative bias parameters, we present our results using three fitting strategies, each using three different values of $k_{\max}$, in this section and use the final strategy as a baseline for the remainder of the paper. All results shown in the rest of this paper are averaged over the four equilateral PNG simulations used in this work, described in section \ref{sec:simulations_and_fitting}.

\begin{figure*}
\centering
\begin{subfigure}{0.5\textwidth}
  \centering
  \includegraphics[width=\linewidth]{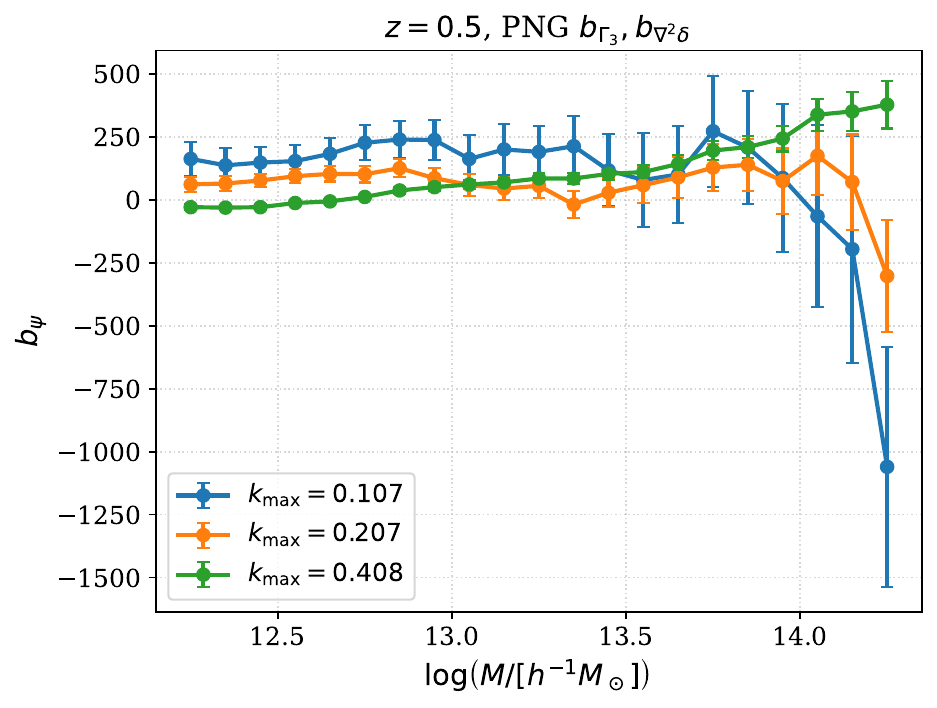}
\end{subfigure}%
\hfill
\begin{subfigure}{0.5\textwidth}
  \centering
  \includegraphics[width=\linewidth]{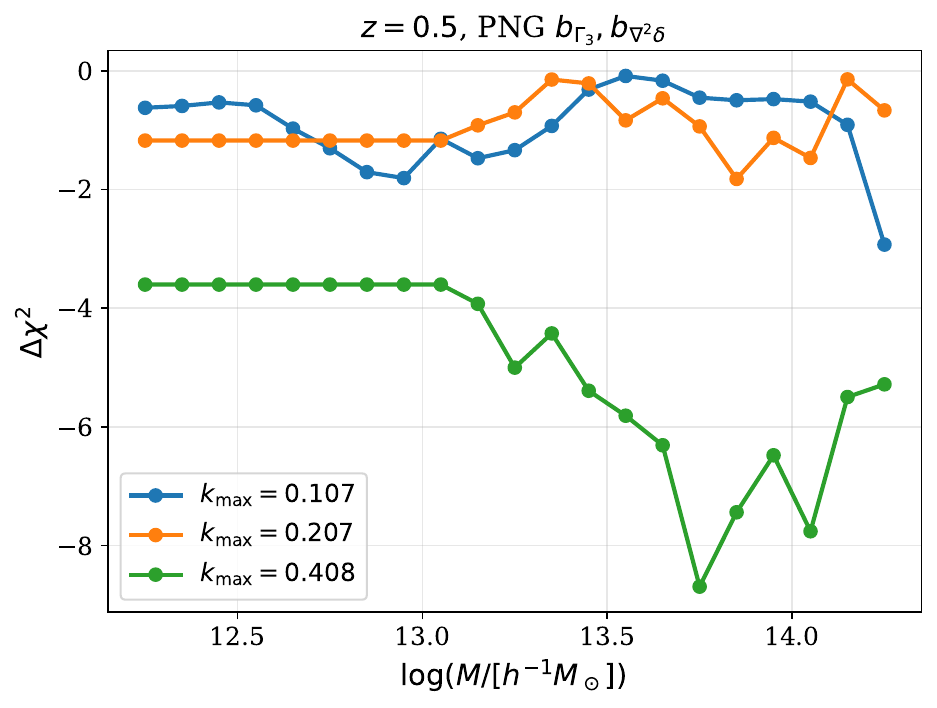}
\end{subfigure}
\caption{\textbf{Left Panel:} Field-level fit values of PNG-induced bias parameter $b_{\psi}$ from our non-Gaussian simulations as a function of halo mass for $z = 0.5$. 
Here, we fit 
$b_1,b_\psi$,
$b_{\Gamma_3}$ and $b_{\nabla^2 \delta}$ using $\beta_1$ from simulations with PNG using eq.~\eqref{eq:final_model}. The curves correspond to different $k_{\max}$ used in fitting. 
\textbf{Right Panel:} Detectability of $b_{\psi}$ as a function of halo mass, measured using the difference in chi-squared statistic $\Delta\chi^2 = \chi^2(b_\psi \neq 0) - \chi^2(b_\psi = 0)$. Increasing $k_{\max}$ increases the sensitivity to $b_{\psi}$ in all mass bins.
For mass bins with $\log M_h \lesssim 12.8\,h^{-1}M_{\odot}$, the number of halos in the Quijote \textsc{Rockstar} suite is not converged. To avoid unreliable noise estimates, we assign these bins a conservative covariance as measured in the first converged bin, $\log M_h$ $\in$ [12.8, 13.3] $h^{-1}M_{\odot}$. This results in the nearly constant $\Delta\chi^2$ plateau visible at low masses.}
\label{fig:bpng_vs_bin_NGbG3}
\end{figure*}

First, we try fitting everything ($b_1, b_{\psi}$ as well as the bias parameters $b_{\Gamma_3}$ and $b_{\nabla^2\delta}$) to the transfer functions obtained from the PNG simulations using eq.~\eqref{eq:final_model}. Our measured $b_{\psi}$ values as a function of halo mass for $z = 0.5$ are shown in the left panel of Figure \ref{fig:bpng_vs_bin_NGbG3}. The three curves correspond to fits with increasing $k_{\max}$ values, to show how the inferred $b_{\psi}$ value changes as more small-scale information is included. To study the sensitivity of our model to $b_{\psi}$ at the field level, we show the difference in $\chi^2$ statistic between our fits with and without $b_{\psi}$, $\Delta\chi^2 = \chi^2(b_\psi \neq 0) - \chi^2(b_\psi = 0)$, in right panel of Figure \ref{fig:bpng_vs_bin_NGbG3}. Overall, we can see that including more small scale information increases our sensitivity to the PNG-induced bias. However, the fits are consistent with $b_{\psi}\approx 0$ for all values of $k_{\max}$. This is reflected in $\Delta\chi^2$ curves in the right panel, where we do not see a statistically significant detection of nonzero $b_{\psi}$ for any curve. 
When both $b_{\Gamma_3}$ and $b_{\nabla^2\delta}$ are free, essentially all of the $b_\psi$ signal is washed out. This shows that $b_\psi$ is degenerate with these parameters - for instance, $b_{\nabla^2\delta}$ captures the $k^2$ part of the PNG-induced scale dependence \cite{Desjacques:2016bnm,Assassi:2015jqa}. As a result, the fits remain consistent with $b_{\psi}\approx 0$ for all $k_{\max}$ values, and $\Delta\chi^2$ is not significant.

\begin{figure*}
\centering
\begin{subfigure}{0.5\textwidth}
  \centering
  \includegraphics[width=\linewidth]{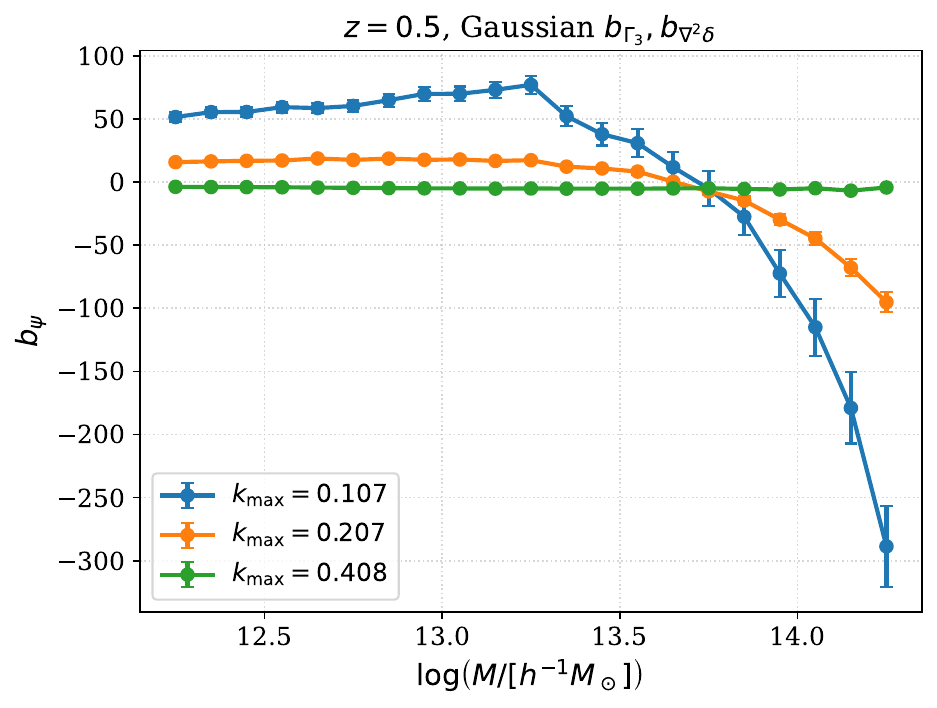}
\end{subfigure}%
\hfill
\begin{subfigure}{0.5\textwidth}
  \centering
  \includegraphics[width=\linewidth]{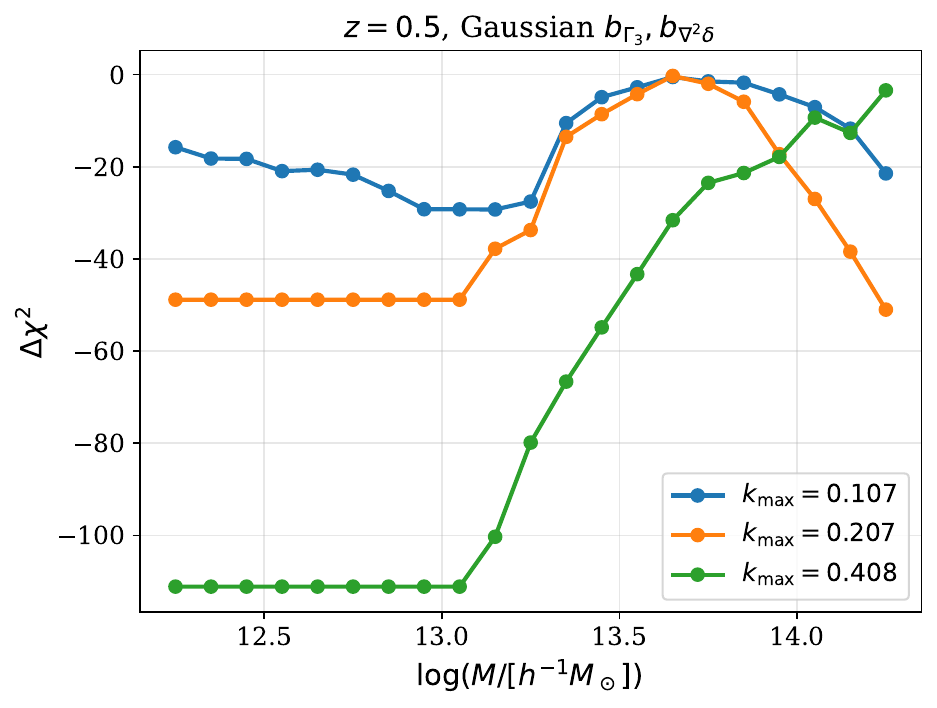}
\end{subfigure}
\caption{Same as Fig. \ref{fig:bpng_vs_bin_NGbG3}, but here we fix $b_{\Gamma_3}$ and $b_{\nabla^2 \delta}$ from Gaussian simulations using eq. \ref{eq:beta_1_G} when performing our fit.
As in Fig. \ref{fig:bpng_vs_bin_NGbG3}, the $\Delta\chi^2$ plateau at low masses is due to our conservative choice of a fixed constant covariance for these low mass bins.
}
\label{fig:bpng_vs_bin_G}
\end{figure*}

To get a nonzero measurement of $b_{\psi}$ by breaking this degeneracy and reducing the number of free parameters, we next fix $b_{\Gamma_3}$ and $b_{\nabla^2 \delta}$ to their values measured from Gaussian simulations (using eq.~\ref{eq:beta_1_G}), leaving only $b_1$ and $b_{\psi}$ free in the fits to $\beta_1$.
Figure \ref{fig:bpng_vs_bin_G} shows our results using this fitting strategy. This fitting strategy pins down $b_\psi$ more tightly, and the sensitivity increases with higher $k_{\max}$. However, because $b_{\nabla^2\delta}$ lies along a strong degeneracy direction with $b_{\psi}$, fixing it to the Gaussian value forces the fit to compensate by changing $b_\psi$ and makes the fits sensitive to the $k_{\max}$ used. In the left panel, we see that the inferred trend in $b_{\psi}$ is sensitive to the choice of $k_{\max}$. At low $k_{\max}$, the fits prefer stronger mass dependence and significant negative values at high mass. At high $k_{\max}$, the fitted $b_{\psi}$ is nearly flat and close to 0, with smaller uncertainties compared to the lower $k_{\max}$ fits. Thus, this fitting strategy leads to biased and inconsistent estimates of $b_{\psi}$, which can also be observed in the $\Delta\chi^2$ curves as a function of $k_{\max}$; the fits do not behave coherently, indicating that the model is mis-specified. 

Given the strong degeneracy between $b_{\nabla^2\delta}$ and $b_{\psi}$, we know that $b_{\nabla^2\delta}$ should not be fixed in order to get robust measurements. Instead, we can try to fix only $b_{\Gamma_3}$. So, in the third approach to reduce the number of free parameters while robustly measuring $b_{\psi}$, we used an empirical linear fit between $b_{\Gamma_3}$ and $b_{\mathcal{G}_2}$ similar to the one measured across 10,500 BOSS-like HOD galaxy mocks with a field-level EFT pipeline in \cite{Ivanov:2024hgq}: $b_{\Gamma_3} = -2.90\, b_{\mathcal{G}_2} - 0.13$. 

\begin{figure*}
\centering
\begin{subfigure}{0.5\textwidth}
  \centering
  \includegraphics[width=\linewidth]{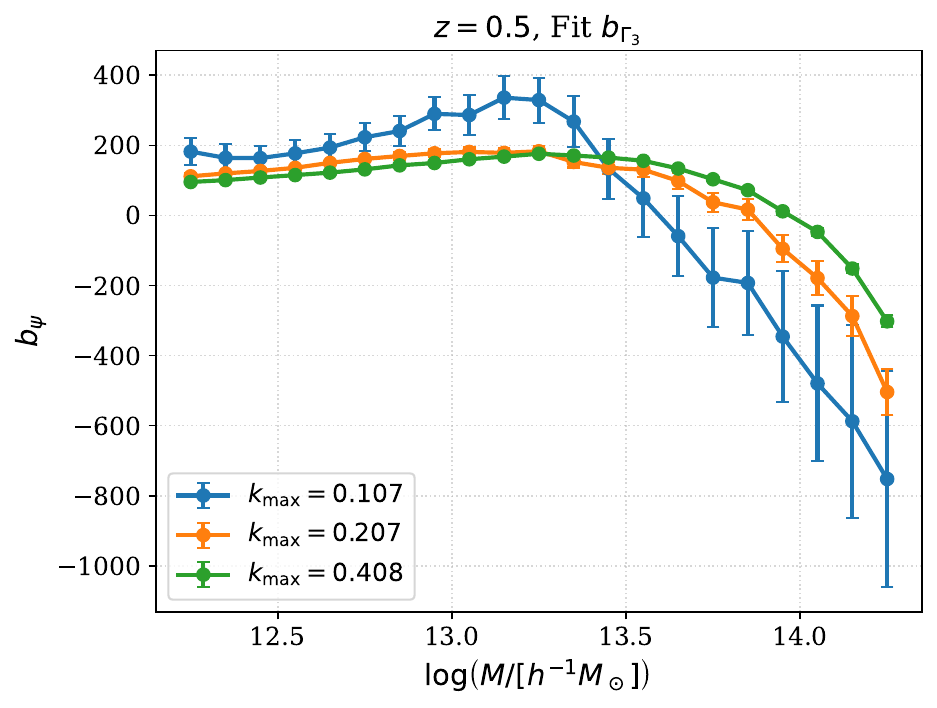}
\end{subfigure}%
\hfill
\begin{subfigure}{0.5\textwidth}
  \centering
  \includegraphics[width=\linewidth]{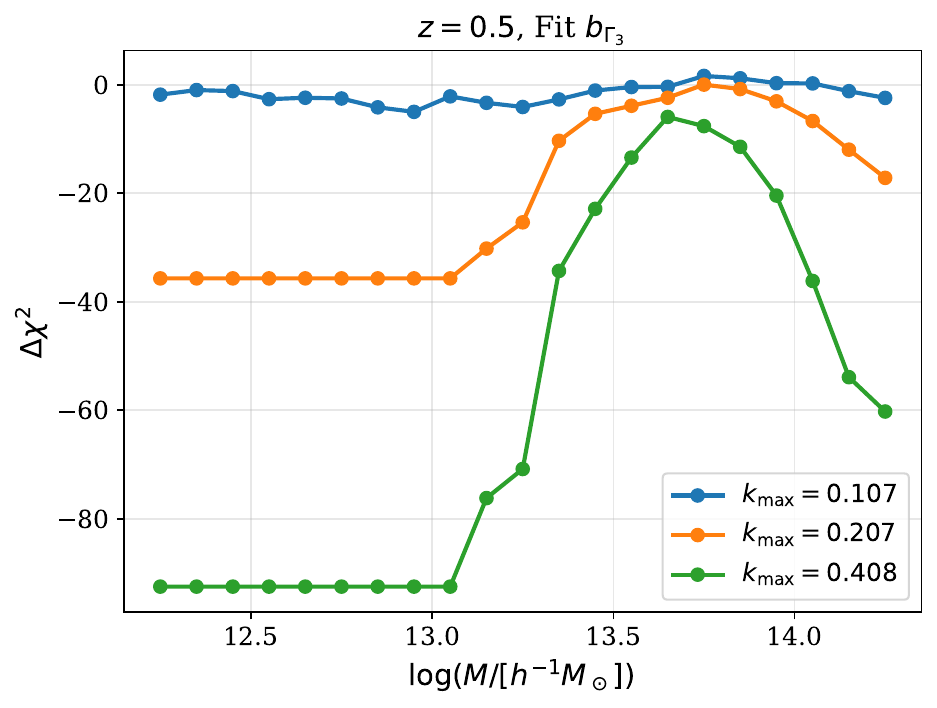}
\end{subfigure}
\caption{Same as Fig. \ref{fig:bpng_vs_bin_G}, but we use an empirical relation $b_{\Gamma_3} = -2.90\, b_{\mathcal{G}_2} - 0.13$ as in~\cite{Ivanov:2024hgq} and fit $b_1,b_\psi, b_{\nabla^2 \delta}$ from $\beta_1$. 
}
\label{fig:bpng_vs_bin_baseline}
\end{figure*}

In Figure \ref{fig:bpng_vs_bin_baseline} we show our fit values of $b_{\psi}$ when fixing $b_{\Gamma_3}$ using this relation and fitting $b_1, b_{\nabla^2 \delta}, b_{\psi}$. The resulting $b_{\psi}$ measurements are stable across $k_{\max}$ values, and we see a weak detection at large scales, and increasingly nonzero $b_{\psi}$ as more small-scale information is included.
At the lowest cutoff $k_{\max}=0.107~h/$Mpc, the fitted $b_{\psi}$ values are relatively small in magnitude, especially for low-mass halos. For the lowest mass halos, the fit is consistent with $b_{\psi}\approx 0$ using this $k_{\max}$. For the more massive halos, however, we measure a non-zero $b_{\psi}$ at this cutoff scale. This is reflected in $\Delta\chi^2$ curve for $k_{\max}=0.107$ in the right panel, which dips below zero at high masses. At this large-scale cut, a non-zero PNG bias is only weakly preferred except in the case of very biased tracers. This is expected because equilateral PNG induces only a weak scale-dependent effect on large scales, so we need very biased halos to notice such small enhancements. As we extend to smaller scales, however, the measured $b_{\psi}$ values become non-zero for most mass bins, and the error bars shrink. The improvement in fit sensitivity is correspondingly large: the right panel shows $\Delta\chi^2 < 0$ for essentially all masses using higher $k_{\max}$. 

As an additional test, 
we apply this fitting strategy to the 
Gaussian simulations, and find 
that our measurements of $b_\psi$
are consistent with zero within $2\sigma$
for all halo masses. This test is described 
in Appendix~\ref{app:gaussfit}.

For the remainder of the paper, we use the strategy based on the empirical
$b_{\mathcal{G}_2}-b_{\Gamma_3}$
relation 
as our baseline when fitting for $b_{\psi}$.

\begin{figure*}
\centering
\begin{subfigure}{0.5\textwidth}
  \centering
  \includegraphics[width=\linewidth]{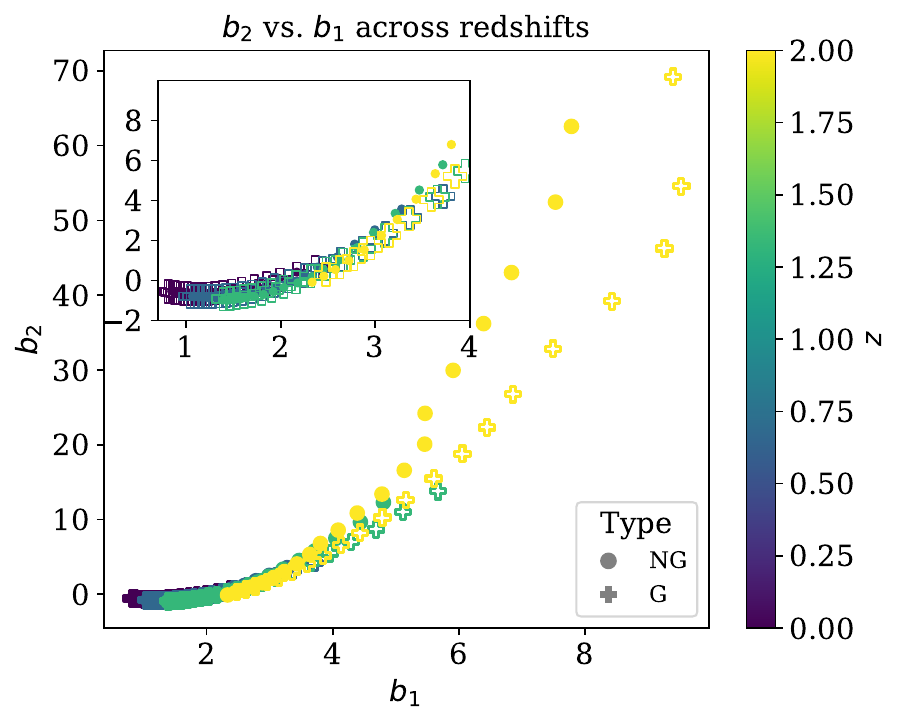}
\end{subfigure}%
\hfill
\begin{subfigure}{0.5\textwidth}
  \centering
  \includegraphics[width=\linewidth]{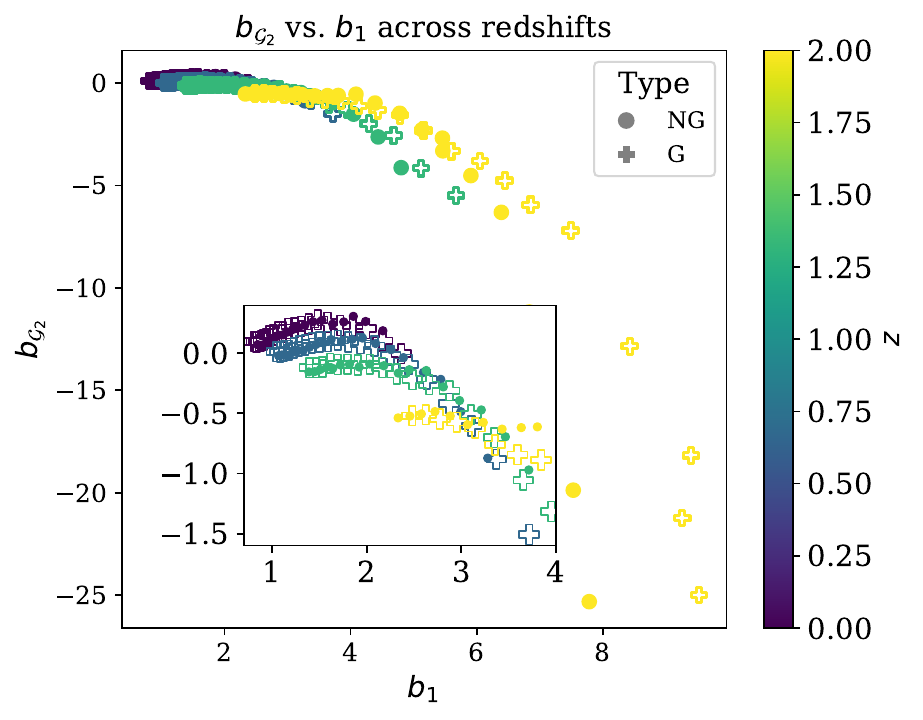}
\end{subfigure}
\caption{Relations between higher-order halo bias parameters and the linear bias $b_1$ across all redshifts and mass bins for both Gaussian (G, pluses) and non-Gaussian (NG, circles) simulations. \textbf{Left Panel:} quadratic bias $b_2$ versus $b_1$, \textbf{Right Panel:} tidal bias $b_{\mathcal{G}_2}$ versus $b_1$. The color scales encode redshift for $z \in [0.0, 0.5, 1.0, 2.0]$. The Gaussian and PNG values follow nearly identical trends, with pluses mostly overlapping with the circles, indicating that PNG does not significantly alter the relationships among the bias parameters. }
\label{fig:b2bG2_vs_b1_baseline}
\end{figure*}

To study the impact of PNG on the relations among bias parameters, we show the $b_2 - b_1$ and $b_{\mathcal{G}_2} - b_1$ trends for all halo mass bins and redshifts, with points colored by redshift, in Figure \ref{fig:b2bG2_vs_b1_baseline}. The tight correlations indicate that both $b_2$ and $b_{\mathcal{G}_2}$ are well described as functions of $b_1$. Importantly, these relations are not significantly modified with PNG, as can be seen from the similar trends that the Gaussian and non-Gaussian samples follow across redshifts. This suggests the validity of our use of Gaussian relations between bias parameters to break degeneracies when fitting $b_\psi$ from PNG simulations.

\subsection{Redshift dependence}
\label{sec:other_redshift_results}

\begin{figure}[htbp]
    \centering
    \includegraphics[width=
    \columnwidth]{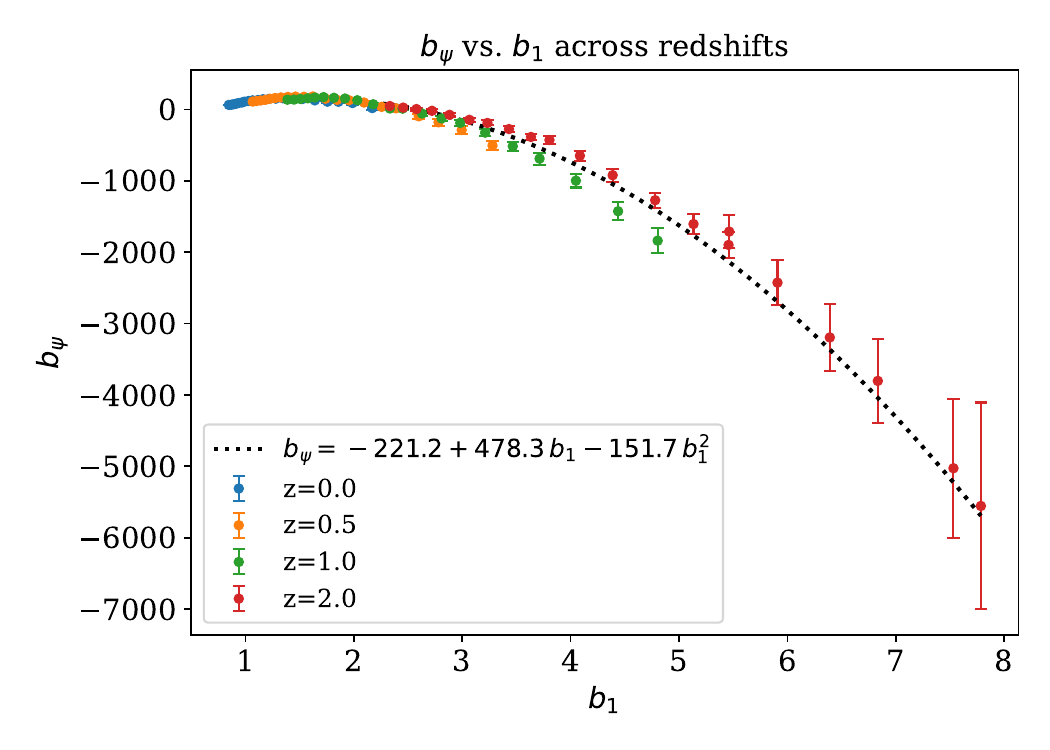}
    \caption{Our measured $b_{\psi}$ values versus the linear bias $b_1$ across multiple redshifts. Each point for a given redshift corresponds to $\log_{10}(M/h^{-1}M_\odot) \in [12.0,12.5], [12.1,12.6], \dots, [14.0,14.5]$. Different colors correspond to the different redshifts used in this work. We see agreement with the PBS expectation that the magnitude of $b_{\psi}$ grows with halo bias.}
\label{fig:bPNGvsb1}
\end{figure}

In Figure \ref{fig:bPNGvsb1}, we present a comparison of the field-level fits of $b_\psi$ at multiple redshifts as a function of the linear bias $b_1$, which is an increasing, monotonic function of halo mass. Hence, each successive point for a given redshift corresponds to an increasing halo mass bin with $\log_{10}(M/h^{-1}M_\odot) \in [12.0,12.5], [12.1,12.6], \dots, [14.0,14.5]$. We obtain our $b_\psi$ values using our baseline fitting strategy with a $k_{\rm max}=0.2~h$Mpc$^{-1}$. 

We expect that halos with larger $b_1$ (i.e. rarer, more massive or higher redshift halos) should show stronger PNG response in $b_{\psi}$. This is precisely what we see in the figure. At low values of $b_1$, $b_{\psi}$ is indeed close to 0. Moreover, the fits reveal a coherent negative trend of $b_\psi$ with increasing halo mass. The effect is weak at low masses but grows in amplitude for higher-mass halos, with the steepest decline at higher redshifts. This is qualitatively consistent with the expectation that rarer, more biased halos show the strongest response to PNG-induced mode coupling.

Looking across redshifts, we also find a consistent picture of redshift evolution of $b_{\psi}$. At $z=0.0$, the signal is weak, with $b_{\psi}$ values clustering near zero. By $z=1.0$, a negative trend becomes more apparent, with amplitudes reaching $\approx$ -2000 for the most massive bins. At $z=2.0$, $b_{\psi}$ values plunge to $\approx -5000$ at high mass. These redshift trends are consistent with the growth of halo bias at early times as halos become rarer. 

Figure \ref{fig:bPNGvsb1} offers another consistency check. Notably, measurements from all four redshifts ($z=0, 0.5, 1.0, 2.0$) fall along the same trend, hence we reproduce similar values of $b_{\psi}$ for similar values of $b_1$ across different mass ranges and cosmic epochs. 
To quantify this trend, we performed a global fit to all the measured values of $b_\psi$ across redshifts and mass bins using a simple quadratic relation,
finding 
\be 
b_\psi = -221+478 b_1-152 b_1^2\,.
\ee 

The resulting fit is shown as the black dotted curve in Figure \ref{fig:bPNGvsb1} and provides a compact description of our measurements, capturing both the overall negative scaling of $b_\psi$ with $b_1$ and the increasing magnitude of this effect for highly biased halos.

Finally, 
let us note that the above relation can be 
re-written 
as a dependence 
of $b_{\psi}$
on the generalized $\fnl$-dependent peak height, which will display its implicit
$\fnl$ dependence. 
However, since this 
dependence can be fully absorbed 
into $b_1$,
the final prior
is expected to be
$\fnl$-independent
to a first approximation,
which will be adequate 
for experimentally
relevant values 
$\fnl\lesssim 300$~\cite{Ivanov:2024hgq}.

\subsection{Comparison with Peak-Background Split Predictions}
\label{sec:PBS}

Let us compare our measurements
now with the 
predictions of the 
phenomenological
peak-background split model 
(PBS).

The predictions from PBS cannot be directly compared against our measurements because of their cutoff dependence. In general, the EFT bias parameters are 
scheme and scale-dependent, and thus their measured values depend on the cutoff scale used within a given framework. A physically measurable quantity is the halo-matter cross power spectrum. Hence, this must be independent of the cutoff scales and can be used to relate measurements from different frameworks. 

The leading PNG one-loop halo-matter cross spectrum is given by
\begin{align}
    P_{hm} \supset b_1 P_{11} + f_{\rm NL} b_\psi \frac{(k/k_{\rm NL})^2}{\mathcal{M}(k)} P_{11} + P_{hm, 12}(k) ,
\end{align}
where  $P_{hm,12}$
is given in eq.~\eqref{eq:Phm12}.
In the UV limit $k\ll p$, the leading term from $P_{hm, 12}$ is the one proportional to $b_2$,
\begin{align}
P_{hm,12}^{b_2} 
&= \int_{\vp} \frac{b_2}{2}\, B_{111}\!\big(k,p,|\k-\vp|\big) \nonumber\\
&= \frac{b_2}{2}\,\mathcal{M}(k)\!
   \int_{\vp} \mathcal{M}(p)\,\mathcal{M}(|\k-\vp|)\,
   b_\psi(k,p,|\k-\vp|).
\end{align}
Now, we can compute the UV limit of this term using the squeezed limit of the primordial bispectrum from eq.~\eqref{eq:Squeezed_equil},
\be
\begin{split}
P^{b_2}_{hm,12} 
 = & ~6 b_2\, k^2\, \mathcal{M}(k)\, P_\phi(k)\, f_{\rm NL}\times
 \\
 &\int_{\vp} \mathcal{M}^2(p)\, P_\phi(p)\,
   \frac{ 1-(\hat{\k}\!\cdot\!\hat{\vp})^2}{p^2}\,.
\end{split}
\ee 
Using $\Lambda$ as the cutoff in the momentum integral, doing the angular integral, and using the linear matter power spectrum $P_{11}(p) = \mathcal{M}^2(p)P_\phi(p)$, we get
\begin{align}
P^{b_2}_{hm,12}(k)
&= 12 b_2\, \frac{k^2}{\mathcal{M}(k)}\, P_{11}(k)\, f_{\rm NL}\,
   \int_0^{\Lambda} \frac{dp}{6\pi^2}\, P_{11}(p)\,,
\end{align}
implying the total cross-spectrum 
contribution
\begin{align}
P_{hm}(k) &\supset b_1 P_{11}(k) 
+ f_{\rm NL}\,\frac{(k/k_{\rm NL})^2}{\mathcal{M}(k)}\,P_{11}(k) \nonumber\\
&\quad\times 
\Big[ b_\psi(\Lambda) 
+ 12 b_2 k_{\rm NL}^2 \int_0^{\Lambda} \frac{dp}{6\pi^2}\, P_{11}(p) \Big].
\end{align}
When we measure the PNG effects on the halo distribution, we effectively probe the scale-dependent second term in the above equation. This measurable must be independent of the cutoff scale, for a given redshift $z$, and so we get the following Renormalization Group (RG) equation:
\be 
12 b_2   \sigma^2_v(\Lambda, z)k_{\rm NL}^2+b_{\psi}(\Lambda)
=12 b_2   \sigma^2_v(\Lambda', z)k_{\rm NL}^2+b_{\psi}(\Lambda')\,,
\ee 
where we have defined 
\be
\sigma^2_v(\Lambda, z) =D^2(z)\int_0^{\Lambda} \frac{dp}{6\pi^2 }P_{11}(p)\,.
\ee
The
PBS predictions correspond to the limit of $k \to 0$,  $\Lambda \to 0$~\cite{Desjacques:2016bnm}, while our field-level measurements correspond to  $\Lambda \to \infty$, which matches the scheme used in CLASS-PT
and data analysis e.g.~\cite{Cabass:2022wjy}. Since $\sigma^2_v(0, z) = 0$, we have the following relation between our measurements, $b_\psi(z)$, and PBS predictions, $b_\psi (\Lambda=0, z)$,
\be
\label{eq:rg_eq}
b_{\psi}(\Lambda=0, z) \;=\; b_{\psi}(z) \;+\; 12\, b_2(z)\, \sigma_v^2(\Lambda\to\infty, z)\, k_{\rm NL}^2 \,.
\ee
This is the prediction appropriate for a comparison to PBS.\footnote{A potential source of discrepancy between PBS and 
EFT could be RG effects of redundant cubic operators such as $\psi\delta_1^2$, $\psi\mathcal{G}_2$.
Their contribution is set to zero
in the scheme used in CLASS-PT, 
which might generate additional 
corrections in the 
$\Lambda \to 0$ limit
required by the PBS. We leave this for future work. 
} To evaluate $b_\psi(\Lambda=0, z)$ in practice, we use $b_2$ extracted from the $\beta_2$ fits.

\begin{figure}[htbp]
    \centering
    \includegraphics[width=\columnwidth]{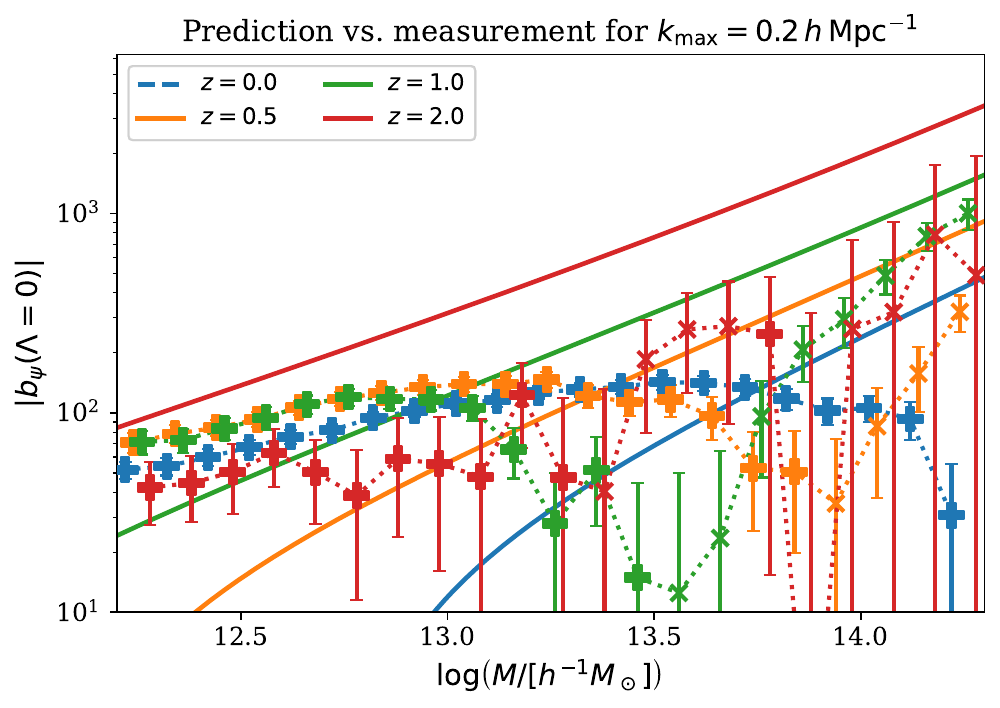}
    \caption{Prediction for $b_{\psi}$ as a function of halo mass and redshift
    from the PBS model (solid curves)~\cite{Schmidt:2010gw} along with our measurements (dotted curves) run to $\Lambda=0$
    using the renormalization group flow~\eqref{eq:rg_eq}.  The marker shape encodes the sign of $b_{\psi}$, with plus symbols corresponding to positive values and crosses to negative values. Across all redshifts, the bias grows with increasing halo mass, and the magnitude of the bias is larger at higher redshifts, in agreement 
    with baseline theoretical expectations.}
\label{fig:fabian_prediction}
\end{figure}

We now compare our field-level measurement of the PNG bias to theoretical predictions from the PBS formalism. In figure \ref{fig:fabian_prediction}, we compare our field-level measurement of $b_{\psi}$ as a function of halo mass and redshift against the theoretical predictions derived from the equilateral PNG model in \cite{Schmidt:2010gw} for the baseline scale cut of $k_{\max} = 0.2 h\,\mathrm{Mpc}^{-1}$. The solid lines show the theoretical predictions for $b_{\psi}$ computed using their expression\footnote{In 
\cite{Schmidt:2010gw}, the PNG correction is written as a scale-dependent shift to the Lagrangian bias, $\Delta b_L^{\mathrm{eq}}(k) \propto 
b_L \delta_c \, k^2 \mathcal{M}^{-1}(k) \sigma_{R,-2}^2/\sigma_R^2$. In our framework, this correction is equivalent to introducing the operator $\psi(\mathbf{k}) \equiv k^2 \phi(\mathbf{k})$ 
with amplitude $b_\psi$. We therefore adopt $b_\psi$ as our parameterization, with Eq.~\eqref{eq:PBS_pred} giving the PBS prediction for its value.} for the correction to the Lagrangian bias, $b_L$, induced by equilateral PNG,
\begin{align}
b_{\psi}
= 6\,b_L\,\delta_c\,\frac{\sigma_{R,-2}^2}{\sigma_R^2} k_{\rm NL}^2 ,
\label{eq:PBS_pred}
\end{align}
where $\delta_c = 1.686$ is the critical overdensity for spherical collapse, and
$\sigma_{R,n}$ are defined using the standard window $W_R(k)$ as
\begin{align}
\sigma_{R,n}^2 \equiv \int \frac{d^3k}{(2\pi)^3}\, k^n\, P_{11}(k)\, W_R^2(k).
\end{align}
For a real-space top-hat of radius $R$, the Fourier-space window is
\begin{align}
W_R(k) \equiv 3\,\frac{\sin(kR) - kR\cos(kR)}{(kR)^3},
\end{align}
where $R$ corresponds to the Lagrangian radius of halos.

Eq. \eqref{eq:PBS_pred} encapsulates a few important scalings. First, the PNG bias contribution is proportional to the Gaussian bias $b_L$ and hence grows for more massive halos. Second, it depends on redshift through $b_L$ and variance $\sigma_{R,n}$, becoming larger at higher redshift for a fixed halo mass. These scalings are clearly visible in the solid lines of Figure \ref{fig:fabian_prediction}.

The dotted lines in the figure show our measurements of $\lvert b_{\psi} \rvert$ using the RG-corrected field-level measurements of the halo-matter linear transfer function $\beta_1(k)$. In both cases, we find that $\lvert b_{\psi} \rvert$ increases with both halo mass and redshift. Qualitatively, we find agreement with the expectation that the bias induced by PNG is enhanced at high masses and redshifts, reflecting the growth of the Lagrangian bias parameter in the Press-Schechter framework \cite{Press:1973iz, Ivanov:2025qie} $b_L \propto \frac{\nu^2 - 1}{\delta_c}, \quad \text{with} \quad \nu = \frac{\delta_c}{\sigma(M, z)}$. 
Quantitatively, we observe 
a tension between our measurements 
and the PBS model predictions.
This is not very surprising 
given that the PBS 
model does not capture 
the complete physics
of halo biasing,
e.g. it does not account 
for tidal interactions. 
The quantitative 
mismatch between  
EFT measurements 
and PBS predictions
highlights the limits 
of empirical structure
formation models,
whose accuracy 
will likely be
insufficient 
for upcoming high
precision galaxy
surveys.

\section{Conclusions}
\label{sec:conclusions}

We have studied the leading order 
scale-dependent 
halo bias due to equilateral primordial
non-Gaussianity from N-body simulations. 
To that end we used the field-level
EFT technique which allowed for a reduction of sampling noise in our measurements. Using an empirical relationship between the cubic
and quadratic tidal biases
suggested by N-body simulations, 
we were able to measure 
the equilateral PNG bias 
across various halo masses and 
redshifts. 

Our work provides the first reliable
estimates of the equilateral PNG bias
parameters, which can be used to set priors
in equilateral PNG analyses of galaxy clustering data, e.g. from the power spectra and bispectra of the DESI data~\cite{Chudaykin:2025aux}.
Conservatively, one could use our measurements to set the width of the 
priors on this parameter to marginalize over. Alternatively, one could use our measurements to compute the analytic 
informative priors along the lines of~\cite{Ivanov:2025qie}, which might lead to better constraints on equilateral PNG.

Going forward, it will be interesting to 
extend our approach to the halo occupation distribution catalogs as in~\cite{Ivanov:2024hgq} 
and full hydrodynamical simulations~\cite{Ivanov:2024dgv}.
It will also be interesting to study 
the higher-order equilateral PNG biases like $b_{\psi\delta}$.
The corrections from these
operators are suppressed
in the power-counting 
relative to $P_{12}$,
but they become important
at the two loop order.\footnote{$\psi\delta$ become
 important only at high orders because the equilateral PNG bispectrum is suppressed in the squeezed limit. This can be contrasted with the local PNG, where contributions from the $\phi\delta$ operator
have the same scaling 
as $P_{12}$, so it 
contributes already
at the one loop. 
}
Another obvious extension of our work is 
a detailed analysis of the other types of PNG: orthogonal 
non-Gaussianity and cosmological collider shapes~\cite{Chen:2009zp,Chen:2009we,Chen:2010xka,Baumann:2011nk,Arkani-Hamed:2015bza,MoradinezhadDizgah:2017szk,MoradinezhadDizgah:2018ssw,MoradinezhadDizgah:2019xun,MoradinezhadDizgah:2020whw,Kumar:2018jxz,Kumar:2019ebj,Reece:2022soh,Cabass:2022oap,Cabass:2024wob}.

From the methodological point of view, 
it will be important to improve
the field-level EFT in order to 
break the degeneracy between 
$b_{\Gamma_3}$ and $b_\psi$.
One way forward would be to include cubic operators like $\Gamma_3$ explicitly
in the field-level forward model. 
This will allow us to 
explicitly validate the assumption
of the $b_{\mathcal{G}_2}-b_{\Gamma_3}$
correlation and in
principle, go beyond 
it. 
Additionally, it will be interesting to 
develop a consistent renormalization group 
treatment of the equilateral PNG parameters
that would clarify their relationship
with EFT loop computation schemes
and redundant operators. 
We leave these  
research 
directions for future investigation. 

\acknowledgments

MI thanks Oliver Philcox for
useful discussions. 
KA acknowledges supports from Fostering Joint International Research (B) under Contract No.21KK0050 and the Japan Society for the Promotion of Science (JSPS) KAKENHI
Grant No.JP24K17056.
Part of the numerical computations were carried out on Cray XD2000 system at Center for Computational Astrophysics, National Astronomical Observatory of Japan.
JMS acknowledges that support for this work was provided by The Brinson Foundation through a Brinson Prize.

\appendix

\section{Validation of the fitting prescription on Gaussian simulations}
\label{app:gaussfit}

\begin{figure}[htbp!]
    \centering
    \includegraphics[width=\columnwidth]{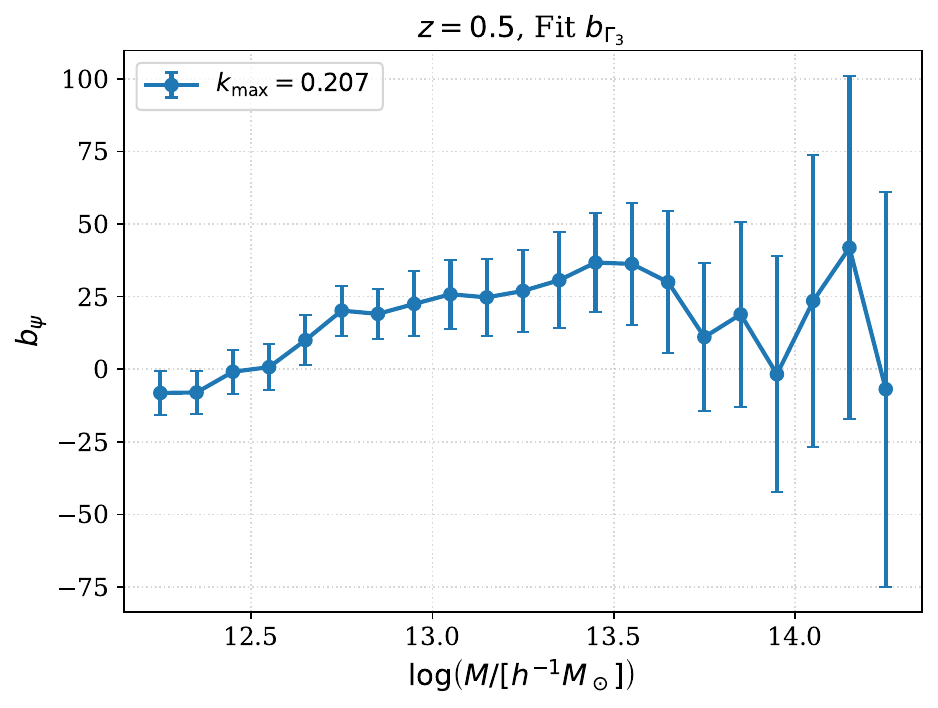}
    \caption{Null test of our baseline $b_\psi$ fitting procudure on Gaussian simulations at $z=0.5$. The fitted values of $b_\psi$ are obtained by applying the same fitting pipeline used for the PNG simulations to a purely Gaussian realization, where the true $b_\psi = 0$. The recovered values are consistent with $b_\psi = 0$, with no systematic trend across halo mass, confirming that our fitting methodology does not introduce spurious PNG bias contributions.}
\label{fig:null_test}
\end{figure}

To validate our fitting strategy and ensure that the extracted PNG bias are not contaminated by numerical artifacts or fitting degeneracies, we apply the full analysis pipeline to a Gaussian simulation in which $f_{\mathrm{NL}} = 0$. In this case, $b_\psi=0$ at all masses. Figure \ref{fig:null_test} shows our recovered $b_\psi$ values as a function of halo mass at $z=0.5$.  For this fit we
used $P_{12}=0$ and set $\fnl=10^3$ to match the main PNG analysis.
The fitted values are consistent with zero within the statistical uncertainties and show no systematic dependence on mass. The figure demonstrates that our fitting procedure is unbiased and robust when applied to Gaussian initial conditions, providing confidence that the non-zero $b_\psi$ values observed in the PNG simulations truly arise from non-Gaussianity within the simulations rather than from the fitting procedure itself.

\bibliographystyle{JHEP}
\bibliography{short.bib} 

\end{document}